\documentclass[cha, preprint]{revtex4-1}
\usepackage{graphicx}
\usepackage{caption}
\usepackage{subcaption}
\usepackage{amsmath}
\usepackage{amsfonts}
\usepackage{color}

\begin{document}

\title{An early warning indicator for atmospheric blocking events using transfer operators}


\author{Alexis Tantet}
\author{Fiona R. van der Burgt}
\author{Henk A. Dijkstra}
\affiliation{Institute for Marine and Atmospheric research Utrecht, Department of Physics and Astronomy, Utrecht University, Utrecht, The Netherlands}

\email{a.j.j.tantet@uu.nl}

\begin{abstract}
The existence of  persistent midlatitude atmospheric flow regimes with time-scales larger than 5-10 days 
and indications of preferred transitions between them motivates to develop  early warning 
indicators for such regime transitions. In this paper, we use a hemispheric barotropic model 
together with estimates of transfer operators on a reduced phase space to develop an 
early warning indicator of the zonal to blocked flow transition in this model. It is shown that, 
the spectrum of the transfer operators can be used to study the slow dynamics of the flow
as well as the non-Markovian character of the reduction.
The slowest motions are thereby found to have time scales 
of three to six weeks and to be associated with meta-stable regimes (and their transitions) which can 
be detected as almost-invariant sets of the transfer operator. From the energy budget of the model, 
we are able to explain the meta-stability of the regimes and the existence of preferred 
transition paths. Even though the model is highly simplified, the skill of the early warning indicator 
is promising, suggesting  that the transfer operator approach can be used in parallel to an 
operational deterministic model	for stochastic prediction or to assess forecast uncertainty. 
\end{abstract}

\maketitle

 \begin{quotation}
	The transfer operator of a model of the atmosphere is used to isolate and detect
	persistent atmospheric regimes  and to define an early warning indicator of transitions 
	between these regimes. The methodology is expected to have  application potential 
	to general dissipative chaotic systems. 
 \end{quotation}

\section{introduction}

The midlatitude atmospheric flow is considered to be a chaotic dynamical system for which 
predictability is limited \citep{Lorenz1963a, Palmer1993}. Although the behavior  of this flow  
is dominated by weather systems on short time scales caused by baroclinic instability, strong 
variability on time scales longer  than  5-10 days, with a predominantly  barotropic structure,  is 
also observed  \citep{James1994}. It has been argued that at least part of the observed 
low-frequency variability can be explained by recurrent and persistent atmospheric regimes 
\citep{Mo1988, Kimoto1993, Smyth1999} such as the North Atlantic Oscillation (NAO) and  
blocking events \citep{Plaut1994, Ghil2002b}. 

Many studies identifying atmospheric regimes use algorithms relying on the recurrence property 
of these regimes such as the k-means \citep{Mo1988, Kondrashov2004a} and the Gaussian mixture 
algorithms \citep{Smyth1999, Smyth2000}. Other studies make use of  persistence properties, for 
example  leading to Hidden Markov  Models \citep{Majda2006, Franzke2008}. Most of these techniques 
rely  on the reduction of the high-dimensional phase space to  a few dimensions.

The existence of weather regimes in General Circulation Models (GCM) and in reanalysis has been 
questioned for some time \citep{Stephenson2004, Christiansen2007, Fereday2008a}. Using the Integrated 
Forecast System (IFS) of the European Centre for Medium-Range Weather Forecasts (talk), it was 
shown \citep{Dawson2012, Jung2005, Dawson2014}  that it was necessary to use a spatial resolution of 
T1279  (16 km),   or to include stochastic parametrizations,  in order for the atmospheric regime behavior 
to occur.  This suggests  that although the atmospheric regimes are large-scale low frequency motions, the 
faster  small-scale motions (either explicitly resolved or included as random perturbations) are  important 
to simulate them. 

The barotropic structure of midlatitude low-frequency variability has motivated early studies using low-order barotropic models. \citet{Charney1979a} have shown that such regimes could manifest themselves in highly truncated spectral barotropic models as stable fixed points representative of different solutions of a standing Rossby wave over topography.  Flow regimes and spontaneous transitions have been observed in laboratory experiments using rotating annulus experiments for a barotropic fluid with topography \citep{Weeks1997} and for a  two-layer shear flow \citep{Williams2003, Williams2004, Williams2005, Williams2008}. In \citep{Weeks1997}, a zonal flow and blocked flow were found for different values of the Rossby number,  and spontaneous transitions between the two were observed for intermediate values of the Rossby number. They suggested 
that these transitions were associated with  the existence of two basins of attraction connected by heteroclinic orbits. 

A scenario of chaotic itinerancy \citep{Kaneko1991a, Itoh1996} permitted by heteroclinic connections is supported by the study of \citet{Crommelin2004a} using a 6-mode barotropic model.  For specific values of the forcing parameter,  the two stable fixed points of the zonal and blocked regimes merge with a periodic orbit (due to  barotropic instability), yielding a heteroclinic connection. Although such a specific situation  is unlikely to exist in the real atmosphere, \citet{Crommelin2003} found evidence of ruins of such a heteroclinic connection in a hemispheric barotropic model with realistic topography and forcing  \citep{Selten1995}, manifested by the  presence of preferred transition paths. 
Regime behavior was also found in more realistic barotropic \citep{Legras1985, Branstator1989, Crommelin2003} and  multilayer quasi-geostrophic models \citep{Itoh1996, Kondrashov2004a}. Because 
these models exhibit chaotic behavior, the regimes are no longer identified by stable fixed points but 
rather as neighborhoods in the phase space where  trajectories tend to persist, motivating their denomination as meta-stable regimes.  

In the laboratory experiments by \citet{Williams2003}, it is the inertia-gravity waves which are responsible for the regime transitions when the flow is baroclinically unstable. Such waves and barotropic disturbances occur in the real atmosphere together with baroclinically unstable synoptic weather systems, so that it is not yet clear if one disturbance is more important than the other in inducing certain regime transitions. All these possible mechanisms suggest, however,  that the variability of the midlatitude atmospheric circulation can be captured by a deterministic model with multiple basins of attraction `forced' by random perturbations representative of high-frequency eddies. 

This multi-scale property of the climate system motivates a stochastic-modeling approach \citep{Gardiner2009} to climate variability \citep{Penland2003}. Stochastic climate modeling often relies on a time-scale separation where the state vector is decomposed into a slow climate component and fast weather fluctuations. These fast fluctuations are  not resolved explicitly but their aggregated effect is represented by a noise term. The system is thus modeled by a Stochastic Differential Equation (SDE) \citep{Gardiner2009} with in general  non-linear deterministic terms and  additive and/or state-dependent noise terms \citep{Majda1999, Majda2001, Franzke2005, Franzke2014}.  When the time-scale separation assumption is violated, the Mori-Zwanzig formalism shows that a non-Markovian term representative of the memory effect induced by past interactions between the resolved and the unresolved variables has to be added \citep{Chorin2009, Darve2009a, Wouters2013a, Kondrashov2014}. Stochastic modeling has been applied to many problems in climate science, such as subgrid-scale parametrization, uncertainty quantification and data assimilation  \cite[]{Palmer2009, Franzke2014}.  

When randomness is present in a dynamical system, whether it is because of uncertainty in the initial state of chaotic systems or because of a stochastic forcing, it is of interest to study the evolution of probability densities in phase space by the flow rather than that of individual trajectories \citep{Lasota1994, Froyland2014}. This evolution is given by the transfer operator whose point spectrum, the Ruelle-Policott resonances \citep{Pollicott1985, Ruelle1986, Ruelle1986a, Butterley2007, Chekroun2014}, give valuable information on the slow dynamics of the system. For mixing dissipative systems, these resonances are associated with a slow correlation decay and the manifestation of meta-stability \citep{Froyland2014}.

The main purpose of the present study is to develop an early warning indicator of transitions between atmospheric flow regimes.  A traditional method that gives an early warning of a sudden transition is the use of the critical slow down of the system when it gets close to a bifurcation point \cite[]{Scheffer2009}. However, in a high-dimensional model such as that used in \citep{Crommelin2003} it is too simplistic to reduce the topology of the system to one or more stable fixed points.  Recently, it was shown that complex networks can reveal information on nearby simple bifurcations in high-dimensional  dynamical systems \citep{VanderMheen2013}.  Near bifurcation points, the topology of the network changes drastically and  early warning indicators for transitions were developed based on these topological changes \citep{Viebahn2014a, Feng2014, Tirabassi2014a}. 

In this study, we base the early warning indicator on the evolution of probability densities with respect to meta-stable regimes in a reduced phase space of the barotropic model used in \citet{Crommelin2003}. To study the slow dynamics in this phase space and to evaluate the effect of memory induced by the reduction, the spectrum of transfer operators estimated for different lags is analyzed. Meta-stable regimes are subsequently detected from the transfer operator at a carefully chosen lag. An early warning indicator of transition to the blocked regime is developed from the transfer operator making use of the existence of preferred transition paths between the regimes.  To test the quality of the early warning indicator, a traditional method employing the Peirce skill score \citep{Peirce1984, Stephenson2000, Thornes2001} is used. Finally, a study  of the energy budget of the barotropic model is performed, where particular attention is given to the conversion 
of mean kinetic energy  to eddy kinetic energy by  Reynolds' stresses, to provide a physical background to the early warning indicator. 

\section{Reduction of the T21-barotropic model}
\label{sec:regime}

\subsection{Model and Data}

Transitions between zonal and blocked regimes of the northern hemisphere atmospheric circulation are
here  investigated using a barotropic model \cite[]{Selten1995, Crommelin2003, Franzke2005}. 
The dimensionless equation of the model, expressed in terms of the streamfunction $\psi$ 
(representing the non-divergent flow) and using the mean radius of the Earth and the inverse of its rotation rate as horizontal and temporal scale,  is given by the barotropic vorticity equation (BVE) 
\begin{eqnarray}
\frac{\partial \nabla^2 \psi}{\partial t} = - \mathcal{J} (\psi , \nabla^2 \psi + f + h_b ) 
- k_1 \nabla^2 \psi + k_2 \nabla^8 \psi + \nabla^2 \psi^*,
\label{eq:bve}
\end{eqnarray}
where $\mathcal{J}$ denotes the Jacobian operator, $f$ the Coriolis parameter, $h_b$ the scaled orography, $k_1$ the Ekman damping coefficient, $k_2$ the coefficient of scale-selective damping and $\nabla^2 \psi^*$ the prescribed vorticity forcing. The 
non-dimensional orography $h_b$ is related to the one of the real Northern 
Hemisphere $h'_b$ by
\begin{eqnarray}
	h_b = 2  A_0 \frac{h'_b}{H} ~ \sin \phi_0 ,
\end{eqnarray}
where $\phi_0 = 45^\circ$N, $A_0 = 0.2$ is a factor determining the strength of the 
surface winds that blow across  the topography, and $H$ is a scale height of 10 km.

The BVE is projected onto spherical harmonics, triangularly truncated at the $21^{st}$ mode (T21). The spherical harmonic coefficients are chosen such that the model is hemispheric with no flow across the equator, resulting in a system of $231$ Ordinary Differential Equations (ODE), which are integrated using a fourth order Runge-Kutta numerical scheme. Following \cite[]{Selten1995}, the Ekman damping time scale and the scale-selective damping time scale were chosen as $15$ days and $3$ days (for wavenumber 21),  respectively,  so as to adequately reproduce  the observed mean and variance of the 500hPa Northern-Hemisphere 10-day mean relative vorticity. The vorticity forcing $\nabla^2 \psi^*$ is calculated from ECMWF reanalysis data of wintertime 500hPa relative vorticity from 1981 to 1991, in order for the first two moments of the simulated relative vorticity to be as close as possible to the observed relative vorticity. The term  $\nabla^2 \psi^*$ is calculated \citep{Roads1987a, Selten1995} according to
\begin{eqnarray}
\nabla^2 \psi^* = \mathcal{J} (\psi_{cl} , \nabla^2 \psi_{cl} + f + h_b ) 
  + k_1 \nabla^2 \psi _{cl} - k_2 \nabla^8 \psi_{cl} + \overline{\mathcal{J} (\psi' , \nabla^2 \psi')},
\label{eq:forc}
\end{eqnarray}
where $\psi_{cl}$ is the mean of the observed streamfunction. The quantity $\psi'$ is the deviation of the 10-day running mean
observed streamfunction from $\psi_{cl}$.  The present study relies on a 500,000-day-long simulation using an integration time-step of 30 minutes, with daily output and a spin-up of $5,000$ days removed.

\subsection{Phase-space reduction}
\label{sec:reducedSpace}

\begin{figure}
	\includegraphics[width=8.5cm]{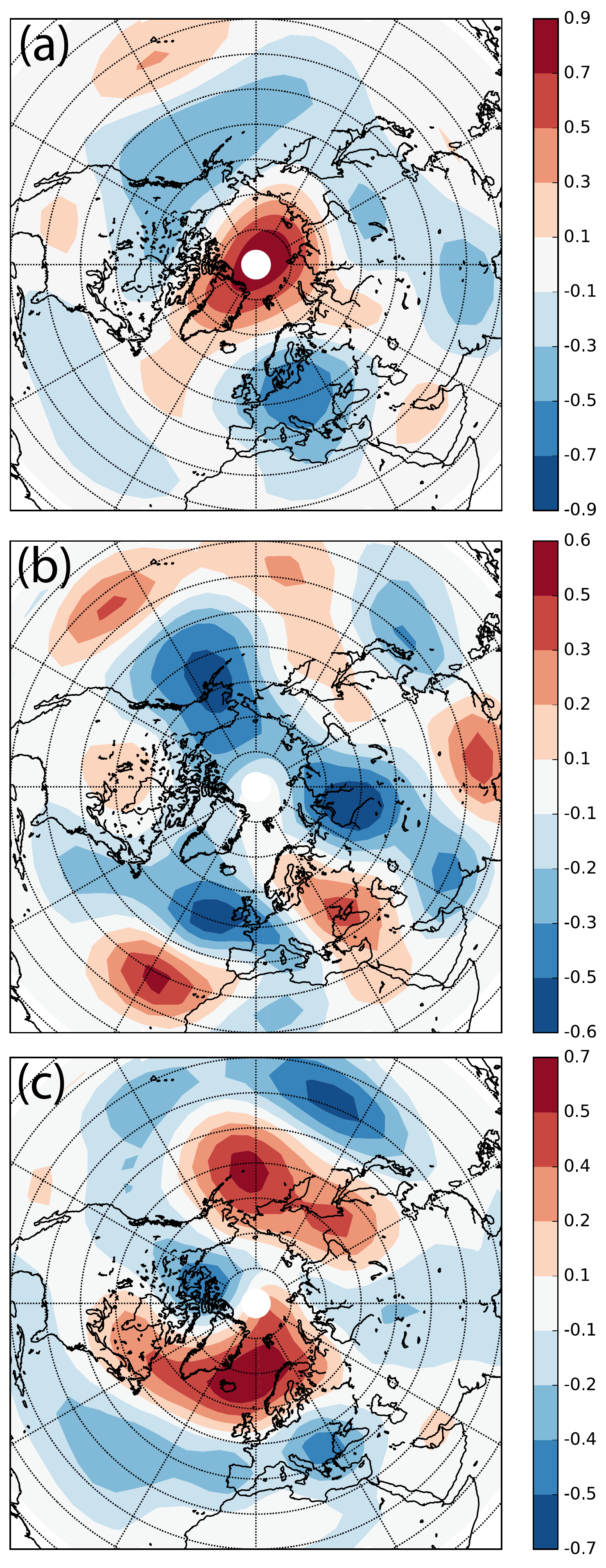}	
	\caption{Dimensionless streamfunction of the three leading EOFs of the barotropic model. 
	EOF 1 (a), EOF 2 (b) and EOF 3 (c) that explain 21.4\%, 8.5\% and 6.8\% of the variance,  respectively. } 
	\label{fig:EOF}
\end{figure}

In order to investigate the presence of meta-stable regimes and preferred transition paths, it is important to define a proper reduction of the $231$-dimensional phase space. This is usually done by projecting the state vector on a low-dimensional basis of orthogonal vectors such that the variance of the projected state vector is maximized. In practice, this can be achieved by an Empirical Orthogonal Function \citep[EOF, e.g ][]{VonStorch1999a} analysis of the streamfunction normalized by the kinetic energy norm \citep{Crommelin2003, Franzke2005}.  The streamfunction patterns of the three leading EOFs are represented in  figure \ref{fig:EOF} and 
explain $36.7\%$ of the total variance. The first EOF is related to the Arctic Oscillation (AO),  blocking events 
and   the strength of the polar vortex \citep{Crommelin2003}. As it shows a dipole-like pattern over the Atlantic basin, 
the third EOF has been associated with  the North Atlantic Oscillation (NAO). 

\begin{figure}[t]
	\includegraphics[width=8.5cm]{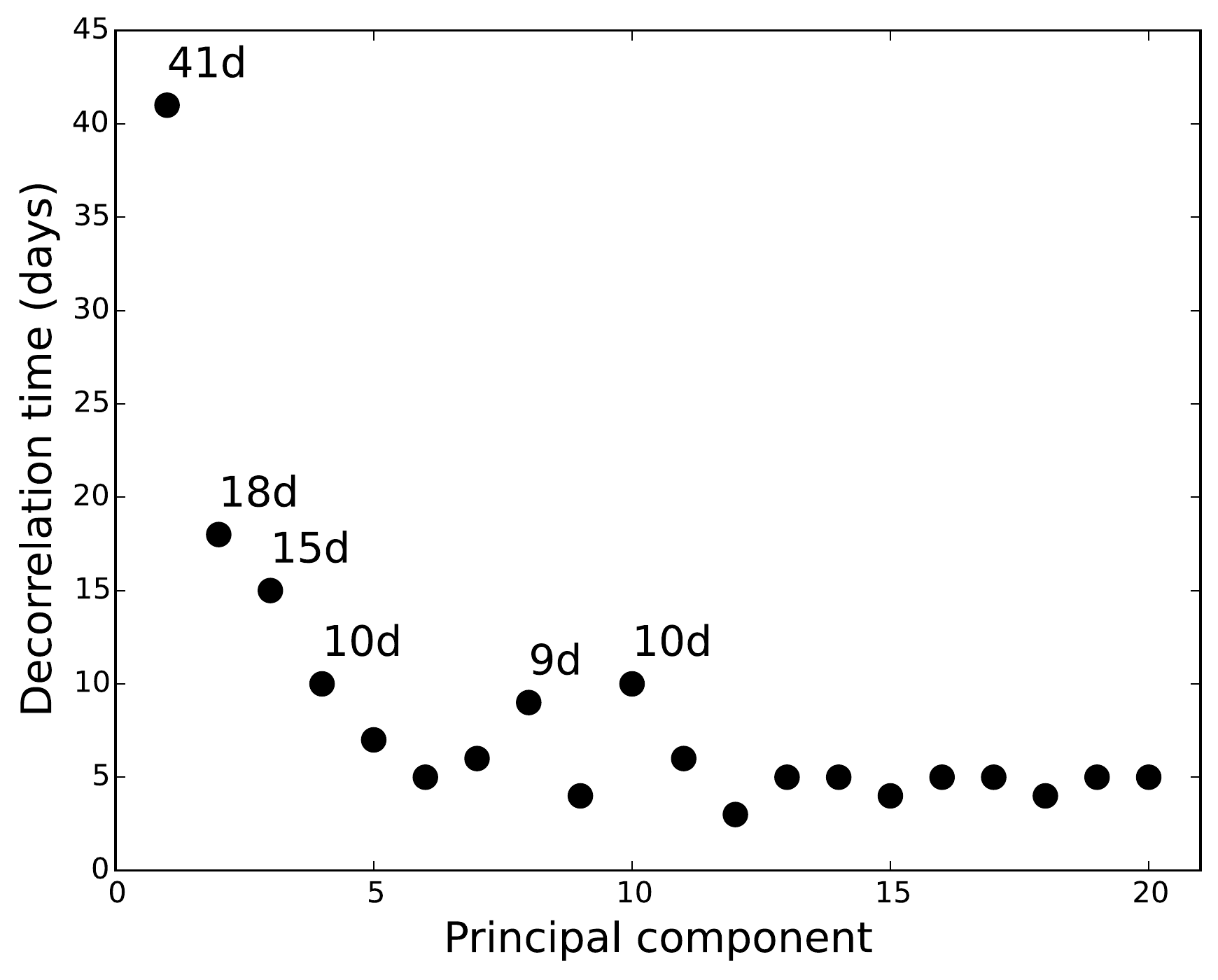}
	\caption{Decorrelation times (in days) of the 20 leading principal components.}
	\label{fig:taud}
\end{figure}

For the purpose of this study, special care must be given to the choice of the EOFs used to define the reduce phase space. Indeed, the presence of meta-stable regimes and the transitions between them should not be hidden by the projection. Furthermore, for the sake of time-scale separation between the motions in the reduced phase space and the unresolved ones, it is important for the principal components of the selected EOFs to show decorrelation times as large as possible compared to the other principal components. These decorrelation times, defined as the time scale after which the autocorrelation function has decayed to $1/e$ of its value at lag 0, are plotted in figure \ref{fig:taud} for the 20 leading principal components. Largest decorrelation times are found for principal components 1, 2 and 3 (41, 18 and 15 days,  respectively), while other principal components have a decorrelation time shorter than 10 days (with many smaller than 5 days). 
Because  \citet{Crommelin2003} has shown that meta-stability was mainly visible on the first principal component ($\text{pc}_1$), while 
transition between the meta-stable regimes occurred  through low and high values of principal component $\text{pc}_3$, we have chosen to define the reduced phase space $Y$ as the $(\text{EOF}_1, \text{EOF}_3)$ plane. The projection  of the state vector 
$\mathbf{x} \in X$ on the reduced phase space $Y$ is then given by the observable 
$\mathbf{y}$ obtained by $h : {\bf x}_t \mapsto {\bf y}_t = (\mathrm{pc}_1(t), 
\mathrm{pc}_3(t))$. 

\begin{figure}[t]
	\includegraphics[width=8.5cm]{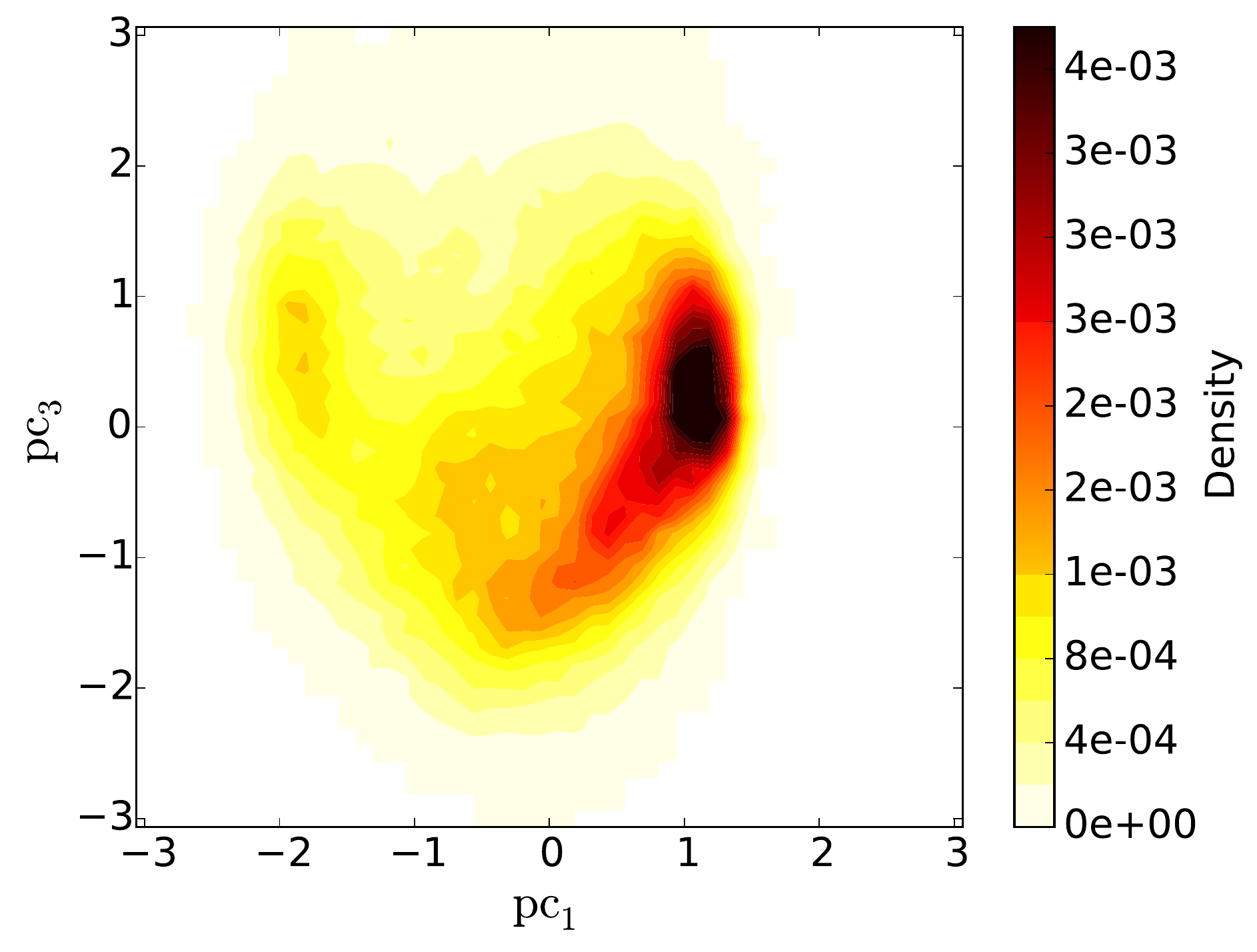}
	\caption{Normalized histogram of the first and third leading principal components normalized by 
	their respective standard deviations and discretized using  $50 \times 50$ grid boxes.}
	\label{fig:histogramPC13}
\end{figure}

The two-dimensional normalized histogram of $\textnormal{pc}_1$ and $\textnormal{pc}_3$ for the 500,000-day-long 
simulation is (cf. figure~\ref{fig:histogramPC13}) similar to figure 7 in \citep{Crommelin2003}. To estimate the histogram, principal components $\text{pc}_1$ and $\text{pc}_3$ were normalized by their respective standard deviations and the $(\text{EOF}_1, \text{EOF}_3)$ plane was discretized into a grid $G$ of $50 \times 50$ boxes spanning a $[-3, 3] \times [-3, 3]$ square. Grid boxes at the boundary are extended so  that all realizations belong to the grid. Furthermore, if a box contains no realization, it is removed from the grid, as it is not likely to support part of the projection of the attractor.  This resulted in a grid of $m = 1577$ boxes containing on average $314$ realizations and such that $80\%$ of the boxes contain at least 20 realizations. The components $H_i$ of the normalized histogram $H$ for grid-box $B_i$ are then calculated as the likelihoods $\hat{\mathbb{P}}(y_t \in B_i) = \#\{y_t \in B_i\} / \#\{y_t \in G\}$, where $\#\{y_t \in B_i\}$ is the number of realizations of the observable $\mathbf{y}$ in box $B_i$ and $\#\{y_t \in G\}$ is the total number of realizations. 

The histogram in figure \ref{fig:histogramPC13} shows  two local maxima indicative of the presence of recurrent or persistent 
regimes. These regimes will be precisely defined in section \ref{sec:metastable}. For now, following \citet{Crommelin2003},
we associate the local maximum for negative pc$_1$ to the blocked regime
(since it corresponds to high pressure over north-western Europe)
and the maximum for positive  values of  $\textnormal{pc}_1$  to the zonal regime.
\citet{Crommelin2003} also showed that the transitions from the zonal to the blocked regime were 
preferably going through negative values of  $\textnormal{pc}_3$, interpreted as the positive phase 
of the NAO.  

The dynamics of the state vector $\mathbf{x}$ of the barotropic model on the phase 
space $X$ are deterministic and Markovian. However, application of the Mori-Zwanzig 
formalism \citep{Chorin2009, Darve2009a, Kondrashov2014} indicates  that in a closed 
model of  the dynamics of the observable $\mathbf{y}$, memory terms accounting for 
past interactions  between the resolved and the unresolved variables have to be included 
together with an, in  general non-white, noise term. Under the assumption that the time 
scales of the resolved variables  are slower by several orders of magnitude than those 
of the unresolved variables, these memory terms can be neglected \citep{Givon2004, 
Majda2001}.  As can be seen in figure \ref{fig:taud},  the decorrelation time of the unresolved 
variables  $\mathrm{pc}_2$ is not smaller that the one of the resolved variables
$\mathrm{pc}_1$ and $\mathrm{pc}_3$ and hence time-scale separation does not 
apply and memory effects should be considered. In sections \ref{sec:transferResonance} and \ref{sec:spectraMetastable}, we show 
how  transfer operator techniques can provide detailed information on this issue. 

\section{Transfer operators, resonances and their estimation}
\label{sec:transferResonance}

The spectrum of transfer operators giving the evolution of densities induced by
the flow can provide valuable information on the time scales of the dynamics; in 
this section we show how these operators can be approximated on a reduced 
phase space. 

\subsection{Transfer operators of dissipative dynamical systems}
\label{sec:general}

We consider an autonomous dissipative dynamical system (DDS) governed by the ODE
\begin{eqnarray}
	\dot{\mathbf{x}} =& F(\mathbf{x}),& \quad \mathbf{x} \in X,
	\label{eq:ODE}\\
	\mathbf{x}(0) =& x_0,& \nonumber
\end{eqnarray}
where $X = \mathbb{R}^d$ is an Euclidean vector space and $F:X \to X$
a smooth vector field with associated flow $\{S_\tau\}_{\tau \ge 0}$.
The evolution of a distribution $f$ of $\mathcal{D}'$ \citep{Rudin1991} induced by the flow $\{S_\tau\}$ is described by a
family of linear operators $\{\mathcal{L}_\tau\}_{\tau \ge 0}$, namely the \emph{Perron-Frobenius operators} or \emph{transfer operators},
such that
\begin{eqnarray}
	<\mathcal{L}_\tau f, g> = <f, g \circ \mathcal{S}_\tau> , \quad \text{for all } g \in \mathcal{C}^\infty,
\end{eqnarray}
where $\circ$ is the composition operator and $<f, g>$ is the action of the distribution $f$ on the smooth function $g$. The family $\{\mathcal{L}_\tau\}_{\tau \ge 0}$ is a \emph{one-parameter semigroup}, that is
\begin{eqnarray}
	\mathcal{L}_{\tau_1 + \tau_2} = \mathcal{L}_{\tau_1} \mathcal{L}_{\tau_2} ~ ; ~  \mathcal{L}_0 =  I
	\label{eq:semigroup}
\end{eqnarray}

While the more intuitive transfer operators acting on (integrable) probability densities \citep[chap 7.4 in][]{Lasota1994}
will be considered in section \ref{sec:estim} to \ref{sec:forecast}, it has been shown that the statistical properties, such as
the decay rate of correlations, of several mixing dissipative systems are related to the point spectrum of transfer operators acting on (larger) spaces of distributions \citep{Gouezel2006, Butterley2007}.
Such statistics are provided by an ergodic invariant measure.
A measure $\mu$ of a measurable space $(X, \mathcal{B}(X))$ is \emph{invariant} under the flow $S_\tau$ if $\mu(S_\tau^{-1}(A)) = \mu(A)$ for all sets $A$ in $\mathcal{B}(X)$. It is \emph{ergodic} if all invariant sets $A$ are trivial subsets of $X$ ($\mu(A) = 0$ or $\mu(A) = 1$). Moreover, a dynamical system is \emph{mixing} if
$\lim_{t \to \infty} \mu(A \cap S_\tau^{-1}(B)) = \mu(A) \mu(B)$, in which case the \emph{correlation function}
\begin{eqnarray}
	\rho_{f, g}(\tau) = \int f g \circ S_\tau - \int f \int g \quad \text{for } f \in L^1, g \in L^\infty,
	\label{eq:corr}
\end{eqnarray}
converges to zero as the lag $\tau$ goes to infinity \citep[chap 4.4 in][]{Lasota1994}.

In the case of a DDS for which the attractor has Lebesgue measure zero, it is important to choose a \emph{physical} invariant measure for which spatial and temporal averages of continuous observables are identical for an initial set of positive Lebesgue measure \citep{Eckmann1985, Young2002}.
It has been shown, for Anosov flows, that a physical measure of Sina\"i-Ruelle-Bowen (SRB) type \citep{Young2002} exists and that an appropriate Banach space $\mathfrak{B}$ can be defined on which the semigroup of transfer operators is bounded and \emph{strongly continuous} \citep[chap. 1.5,][]{Engel2001}, with its point spectrum $P\sigma(\mathcal{L}_\tau)$ located inside the complex unit-disk \citep{Butterley2007}.

Whether the system considered in this study is Anosov is a difficult question. However, it has been observed that many-particle systems with sensitive dependence to initial conditions "behave" as Anosov \citep{Gallavotti1995}. We will follow this so-called \emph{chaotic hypothesis} and assume
that the transfer operators $\{\mathcal{L}_\tau\}_{\tau \ge 0}$ associated with the flow $\{S_\tau\}_{\tau \ge 0}$ constitute a strongly continuous semigroup and that there exists a unique SRB measure (i.e, 1 is the only eigenvalue of the transfer operators located on the unit disk, the system is mixing).

For a bounded and strongly continuous semigroup, there exists an infinitesimal operator $\mathcal{A}$, namely the \emph{generator} of the semigroup, and $u(\tau) =  \mathcal{L}_\tau f$ is the unique solution ($\forall \tau \ge 0$) of the following Cauchy problem \citep[chap. 2.6 in][]{Engel2001}
\begin{eqnarray}
	\dot u(\tau) = \mathcal{A} u(\tau) ~  ; ~ u(0) = f.
	\label{eq:Liouville} 
\end{eqnarray}
For transfer operators, the linear system (\ref{eq:Liouville}) is the \emph{Liouville equation} which replaces the non-linear problem (\ref{eq:ODE}) at the expense of having to deal with a Partial Differential Equation (PDE).

The Spectral Mapping Theorem (SMT) \citep[chap. 4.3.7,][]{Engel2001} allows us to relate the point 
spectrum $P \sigma(\mathcal{A})$ of the generator of a strongly continuous semigroup to 
the point spectrum  $P \sigma(\mathcal{L}_\tau)$ of the semigroup operators, according to 
\begin{eqnarray}
	P\sigma(\mathcal{L}_\tau) \setminus \{0\} = e^{\tau P \sigma(\mathcal{A})}.
\end{eqnarray}
A vector $\phi$ is called an eigenvector associated with the eigenvalue $\lambda$ in $P\sigma(\mathcal{L}_\tau)$
if $\mathcal{L}_\tau \phi = \lambda \phi$. 

The SMT will be used in section \ref{sec:spectra} to associate the spectrum of the transfer operators to motions of different time-scales. To fix the ideas, let us take an eigenvector $\phi$ (possibly a distribution) associated with a non-zero eigenvalue $\lambda$ of $P \sigma(\mathcal{L}_\tau)$. For any observable $g$ (a smooth test function)
\begin{eqnarray}
	<\phi, g \circ S_\tau>	 = & <\mathcal{L}_\tau \phi, g >\\
				=& \lambda <\phi, g >.
	\label{eq:rhophi}
\end{eqnarray}
Applying the SMT, there exists an $\alpha$ in $P \sigma(\mathcal{A})$ such that $\lambda = e^{t \alpha}$, so that (\ref{eq:rhophi}) becomes
\begin{eqnarray}
	<\phi, g \circ S_\tau> = e^{t Re(\alpha)} e^{i t Im(\alpha)} <\phi, g>.
	\label{eq:decay}
\end{eqnarray}
Thus, the term $<\phi, g \circ S_\tau>$ converges exponentially fast to zero at the rate $-Re(\alpha) > 0$.
Loosely speaking, the point spectrum of the generator $\mathcal{A}$ close to the imaginary axis (correspondingly, the point spectrum of $\mathcal{L}_\tau$ close to the unit circle) is responsible for a slow rate of decay of correlations in the direction of their eigenvectors. This point spectrum, called the \emph{Ruelle-Pollicott (RP) resonances} \citep{Pollicott1985, Ruelle1986, Ruelle1986a}, is therefore a candidate for a mathematical explanation of the observed low-frequency variability such as induced by meta-stability. This important fact motivates the estimation of the RP resonances to study the dynamics associated with meta-stable regimes in section \ref{sec:spectraMetastable}. In sections \ref{sec:estim} to \ref{sec:spectraMetastable}, we present how these resonances can be calculated from approximations of the transfer operators on the reduced phase space.

\subsection{Estimation of transfer operators on the reduced phase space}
\label{sec:estim}

For Anosov systems, the stability of the SRB measure \citep{Kifer1986, Dellnitz1999a}, the transfer operators \citep{Blank2001} and the dominant part of their spectra \citep{Keller1998, Baladi1999} to perturbations, has motivated the discrete approximation of transfer operators 
using Ulam-type methods \citep{Dellnitz1999a, Froyland2001, Blank2001, Chekroun2014}.

For low-dimensional systems, \emph{Ulam's} method relies on a Galerkin approximation of the infinite-dimensional transfer operators by finite-dimensional matrices \citep{ulam1964collection, Froyland1998a, Dellnitz1999a}. For this purpose, the phase space is discretized into a grid $G = \{B_i\}_{1 \le i \le m}$ of $m$ grid boxes defining an orthogonal basis of characteristic functions. The transfer operator $\mathcal{L}_\tau$ is then approximated by a time-homogenous Markov chain with transition matrix $P_\tau$ whose elements are the transition probabilities
\begin{eqnarray}
	(P_\tau)_{ij} = \mathbb{P}(x_{t+\tau} \in B_j | x_t \in B_i),
	\label{eq:transProb}
\end{eqnarray}
of a trajectory of $\mathbf{x}$ in box $B_i$ to pass through $B_j$ after a lag $\tau$. These probabilities can be estimated from a long time-series $\{x_t\}_{1 \le t \le N}$ of $\mathbf{x}$ using the Maximum Likelihood Estimator (MLE)
\begin{eqnarray}
	(\hat{P}_{\tau})_{ij} = \hat{\mathbb{P}}(x_{t+\tau} \in B_j | x_t \in B_i) = \frac{\#\{(x_t \in B_i) \wedge (x_{t+\tau} \in B_j)\}}{\#\{x_t \in B_i\}},
	\label{eq:transEstim}
\end{eqnarray}
where $\#\{x_t \in B_i\}$ is the total number of realizations of $\mathbf{x}$ in $B_i$ and $\#\{(x_t \in B_i) \wedge (x_{t+\tau} \in B_j)\}$ is the number of realizations of $\mathbf{x}$ in box $B_i$ ending in $B_j$ $\tau$-time later. As $N$ goes to infinity, $(\hat{P}_{\tau})_{ij}$ converges to $(P_\tau)_{ij}$ with an error of order $\mathcal{O}(N^{-1/2})$ \citep{Billingsley1961}.

Because, the high dimensionality of the phase space of the barotropic model ($d = 231$) prohibits the direct approximation of $\mathcal{L}_\tau$ and  because, as described in section \ref{sec:reducedSpace} and in \citet{Crommelin2003}, the dynamics relevant to transitions between meta-stable regimes predominantly occur in the reduced phase space $Y = (\mathrm{EOF}_1, \mathrm{EOF}_3)$, we instead estimate the transition probabilities of a Markov process associated with the dynamics in the reduced phase space $Y$ \citep{Chekroun2014}, by replacing $\mathbf{x}$ in (\ref{eq:transProb}) and (\ref{eq:transEstim}) by the observable $\mathbf{y} = h(\mathbf{x})$. Markov operators approximated by these transition probabilities cannot be expected to give the transfer of densities in the reduced phase space as induced by the flow $S_\tau$. Nonetheless, it has been shown by \citet[Theorem A]{Chekroun2014}, for systems with a unique SRB measure and a continuous observable $h$, that the transition probabilities in the full phase-space $X$, are related to the transition probabilities of the Markov process on the reduced phase-space $Y$ by
\begin{eqnarray}
	\mathbb{P}(y_{t+\tau} \in B_j | y_t \in B_i) = \mathbb{P}(x_{t+\tau} \in h^{-1}(B_j) | x_t \in h^{-1}(B_i)).
	\label{eq:theoremA}
\end{eqnarray}
Hence  the transition probabilities in $X$ restricted to pairs 
$(h^{-1}(B_i), h^{-1}(B_j))$ are exactly the transition probabilities in $Y$ 
for pairs $(B_i, B_j)$. As a consequence, the estimated transition matrices $\{\hat{P}_{\tau}\}$ contain information on the transfer operators $\{\mathcal{L}_\tau\}$ and their RP resonances as filtered by the observable. 

Accordingly, we have discretized the reduced phase space $Y$ using exactly the same $50\times50$ grid as for the histogram as in section \ref{sec:reducedSpace}. Then, the matrices $\{\hat{P}_\tau\}$ were estimated from the 500,000-day-long time-series of $\mathbf{y}$ and for different lags $\tau$. Importantly, $\{\hat{P}_\tau\}_{\tau \ge 0}$ need not be a semigroup because (i) the partial observation of the system introduces memory effects \citep{Darve2009a, Chorin2009}, (ii) the Galerkin approximation adds numerical diffusion \citep{Froyland2013b} and (iii) the estimation of transition probabilities from a time-series of limited length is prone to sampling errors (cf. section \ref{sec:spectraMetastable} below).

Before to study in detail the spectral properties of the transition matrices, we remark
that these matrices are row stochastic, implying that their eigenvalues satisfy $|\lambda | \le 1$. In our case,
there exists necessarily a unique eigenvalue equal to $1$ because the transition matrices $\{\hat{P}_\tau\}$, by construction from a unique time series, represent irreducible Markov chains \citep{Baladi2000}. The uniqueness of this eigenvalue in turn implies that its associated 
eigenvector $\vec{\pi} = \{\pi_1, ..., \pi_m\}$, or fixed point, is in fact identical to the normalized histogram $H$ of figure \ref{fig:histogramPC13}.

Indeed the following calculation shows that the histogram $H$ is invariant for any $\hat{P}_\tau$
\begin{eqnarray*}
	(H \hat{P}_\tau)_j =& \sum_{k=1}^m H_k (\hat{P}_\tau)_{kj}\\ 
	=& \sum_{k=1}^m \frac{\#\{y_t \in B_k\}}{\#\{y_t \in G\}}
	\times \frac{\#\{(y_t \in B_k) \wedge (y_{t+\tau} \in B_j)\}}{\#\{y_t \in B_k\}}\\
	=& \sum_{k=1}^m \frac{\#\{(y_t \in B_k) \wedge (y_{t+\tau} \in B_j)\}}{\#\{y_t \in G\}}\\
	=& \frac{\#\{y_{t+\tau} \in B_j\}}{\#\{y_t \in G\}}\\
	=& H_j.
\end{eqnarray*}

Thus, each component $\pi_i$ yields the likelihood $\hat{\mathbb{P}}(y_t \in B_i)$ for realizations of the observable to be in $B_i$.
It is straightforward to associate to any m-dimensional vector on the grid $G$ the corresponding density it approximates \citep{Froyland2009}.
Not doing the distinction between a probability vector and the corresponding probability density should thus not lead to any confusion. 

 
\section{Spectral properties, memory and almost-invariant sets}
\label{sec:spectraMetastable}

\subsection{Spectral properties and slow dynamics}
\label{sec:spectra}

We now apply the Spectral Mapping Theorem (SMT) in order to approximate the RP resonances (the dominant eigenvalues of the generator) from the estimated transition matrices $\hat{P}_\tau$, as filtered by the observable $h$. For this purpose, we first solve, for each estimated transition matrix in $\{\hat{P}_\tau\}$, the eigenvalue problem
\begin{eqnarray}
	\vec{e}_i(\tau) \hat{P}_\tau = \lambda_i(\tau) \vec{e}_i(\tau) \quad \text{for } i \in \{1,..., m\},
	\label{eq:eigen}
\end{eqnarray}
where $\vec{e}_i(\tau)$ is the left-eigenvector associated with the $i^{th}$ eigenvalue $\lambda_i(\tau)$ of $\hat{P}_\tau$. The spectrum of  $\hat{P}_\tau$ changes with the lag $\tau$. However, if $\{\hat{P}_\tau\}_{\tau \ge 0}$ was a semigroup with generator $\mathcal{A}_m$, the spectrum $\{\alpha_i\}$ of $\mathcal{A}_m$ would be independent of $\tau$ and applying the SMT would give
\begin{eqnarray}
	r_i = -Re(\alpha_i) = -\frac{1}{\tau} \log{|\lambda_i(\tau)|} \quad \text{for } \tau > 0 \text{ and for } i \in \{1,..., m\},
	\label{eq:rates}
\end{eqnarray}
where $r_i$ would be the exponential rate of decay of correlation of any observable in the reduced phase-space with the eigenvector associated with $\lambda_i$.

However, as noted section \ref{sec:estim}, the transition matrices $\{\hat{P}_\tau\}_{\tau \ge 0}$ do not necessarily inherit from the semigroup property (\ref{eq:semigroup}) of the transfer operators $\{\mathcal{L}_\tau\}_{\tau \ge 0}$. Thus, no generator may exist and the rates $r_i$ in (\ref{eq:rates}) may depend on the lag $\tau$. Nevertheless, calculating the rates $r_i(\tau)$ for each lag allows (i) to give an approximation of the dominant RP resonances with a control on the lag and (ii) to test the semigroup property (\ref{eq:semigroup}).

\begin{figure}
	\centering
	\includegraphics[width=8.5cm]{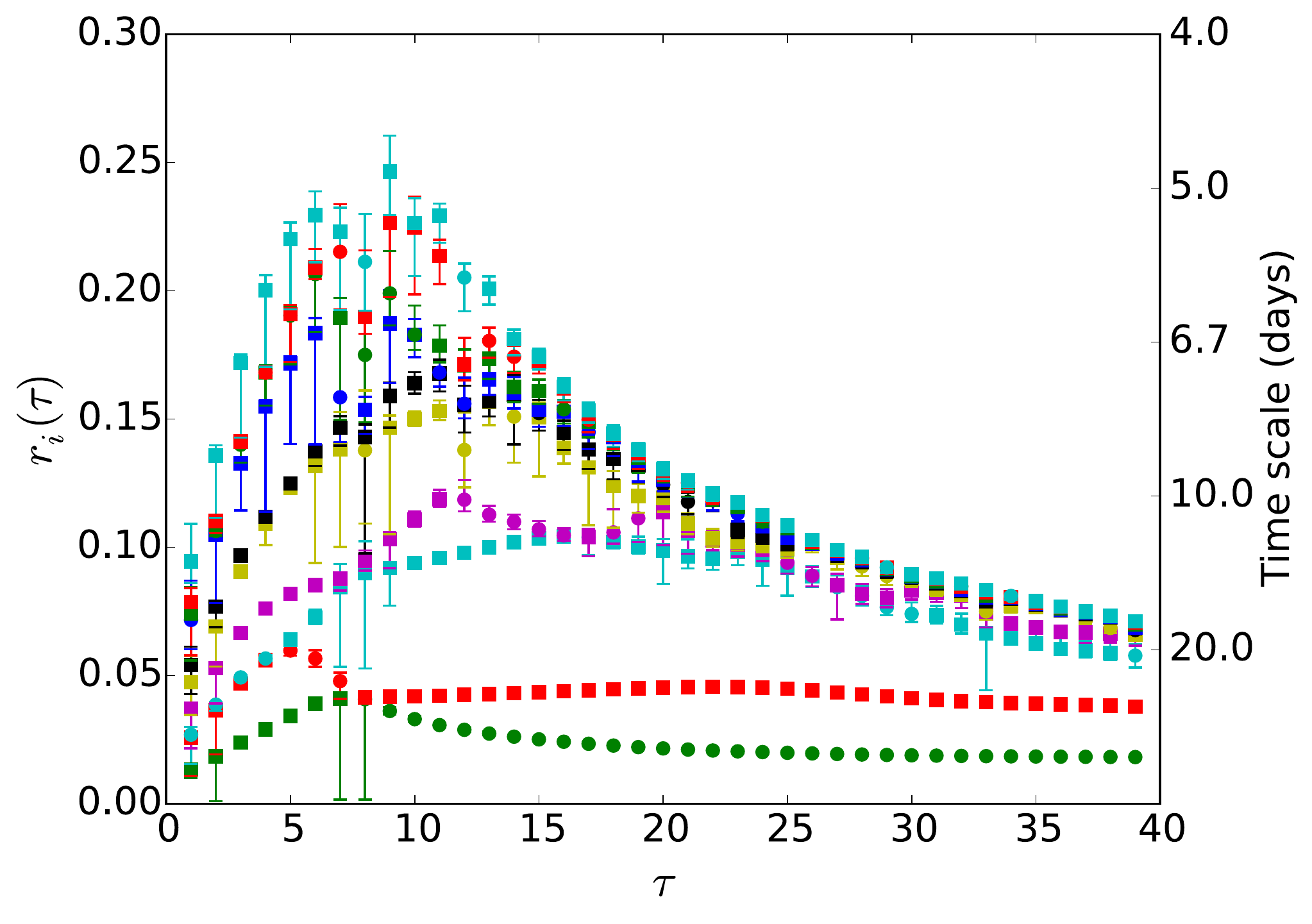}
	\caption{Rates $r_i(\tau)$ corresponding to the 10 leading eigenvalues different from unity of each $\hat{P}_\tau$, with the lag $\tau$ as abscissa and the (cyclic) coloring giving the rank of the rate. A complex pair of eigenvalues is represented by one square for the two conjugates.  The error bars represent $99\%$ confidence intervals estimated from a thousand surrogate transition matrices by applying the bootstrap method described in Appendix \ref{app:A}.}
	\label{fig:spectra}
\end{figure}

For this purpose, we calculated the rates $r_i(\tau)$ by solving the eigenvalue problem (\ref{eq:eigen}) and applying (\ref{eq:rates}) for $\tau$ ranging from 1 to 39 days. The leading rate equals $0$, since it is associated to the unit-eigenvalue. The rates corresponding to the 10 leading eigenvalues different from unity of each $\hat{P}_\tau$ are represented figure \ref{fig:spectra} with the lag $\tau$ as abscissa and the (cyclic) coloring distinguishing the rank of the rate. A complex pair of conjugate eigenvalues is represented by one square for the two conjugates. The error bars represent $99\%$ confidence intervals estimated from a thousand surrogate transition matrices by applying the bootstrap method described in Appendix \ref{app:A}. 

Our first observation concerns the small width of the confidence intervals (some are even hidden by the marker size). These intervals evaluate the robustness of the estimates to the limited length of the time-series. The largest intervals occur when two rates almost overlap, so that one of them may appear or disappear in the surrogates, resulting in a change of rank for all higher-rank rates. In our case, large confidence intervals are thus usually indicative of an uncertainty in the existence of two close but distinct rates or of only one (as for the leading rates at $\tau = 7$ or $8$, for example).
Moreover, Appendix \ref{app:A} shows that at least the three leading rates are very robust to changes in the the grid resolution, ranging from a $10\times10$ grid to a $100\times100$ grid, compared to the original $50\times50$ grid.

Being confident on the robustness of the rates to the sampling as well as to the grid resolution, we now turn to a detailed analysis of the results in  figure \ref{fig:spectra}. First, we can see that the two rates closest to zero, the leading rates in red and green, are well separated from the rest of the spectrum after a lag between 5 and 10 days. One of these rates derives from a real eigenvalue (the circle), the other represents a pair of complex conjugate eigenvalues (the square); they exchange rank at a lag of 8 days. We can calculate an indicative time scale associated with each rate as the inverse of the rate. Whether this time scale corresponds, to a good approximation, to a decorrelation time is not the matter of this study. For a lag of 15 days, the first rate (in green) and the second rate (in red) correspond to a time scale of $40$ and $23$ days respectively. Furthermore, the second rate is separated from the third (in cyan) by a gap of 13 days. This time-scale separation suggests that the reduced dynamics are slowly mixing due to the presence of meta-stable regimes responsible for low-frequency variability \citep{Dellnitz1997a, Froyland2003, Chekroun2014}. This confirms the work on meta-stable atmospheric regimes in this model by \citet{Crommelin2003};  we will see in section \ref{sec:metastable} how such regimes can be more objectively detected.

The second important feature, relevant to the problem of stochastic prediction of section \ref{sec:early}, is the relative constance of the two leading rates for lags larger than $8$ days. We can say that the slow dynamics associated with these rates "behave as Markovian", by which we mean that looking only at these rates, one cannot disprove the semigroup property (\ref{eq:semigroup}), even though the dependance on the lag of the other rates is clearly indicative that $\{\hat{P_\tau}\}_{\tau \ge 0}$ cannot constitute a semigroup. Thus, the two leading rates do not seem to be affected by memory effects due to the partial observation of the system or by estimate errors, since one would not expect them to be constant otherwise.

From the separation of the two leading rates from the other rates as well as their relative independence on the lag $\tau$ for lags larger than $8$ days, we expect $8$ days to be the minimum lag for which the  transition matrix $\hat{P}_{\tau = 8}$ is likely to predominantly resolve the dynamics associated with the meta-stable regimes. Consequently, the following developments will rely mostly on the transition matrix $\hat{P}_{\tau = 8}$. Such strategy for the choice of the lag is similar to the one of \citet{DelSole2000a} and \citet{Berner2005a}, who look directly at the decorrelation rate of their time series to infer for which lag they should estimate the drift and diffusion coefficients of the Fokker-Planck equation they want to approximate. In our case, however, all the rates associated with the dominant eigenvalues of the transfer operators are considered and not only the decorrelation rate of the time series alone. Let us also acknowledge that the SMT has been used in \citet{Crommelin2011} to estimate the spectrum of the generator associated with a Fokker-Planck equation, in \citet{Froyland2013b} to compare approximates of transfer operators and generators of low-dimensional dynamical systems and in \citet{Franzke2008} to test the meta-stability of Hidden Markov Models. However, we do not know other studies using the SMT to test the "Markovianity" of reduced systems, even though the $\tau$-test in \cite{Penland1995} serves a similar purpose in the context of Linear Inverse Modeling.

\subsection{Further validation of the semigroup property}
\label{sec:semigroup}

\begin{figure}
	\centering
	\includegraphics[width=8.5cm]{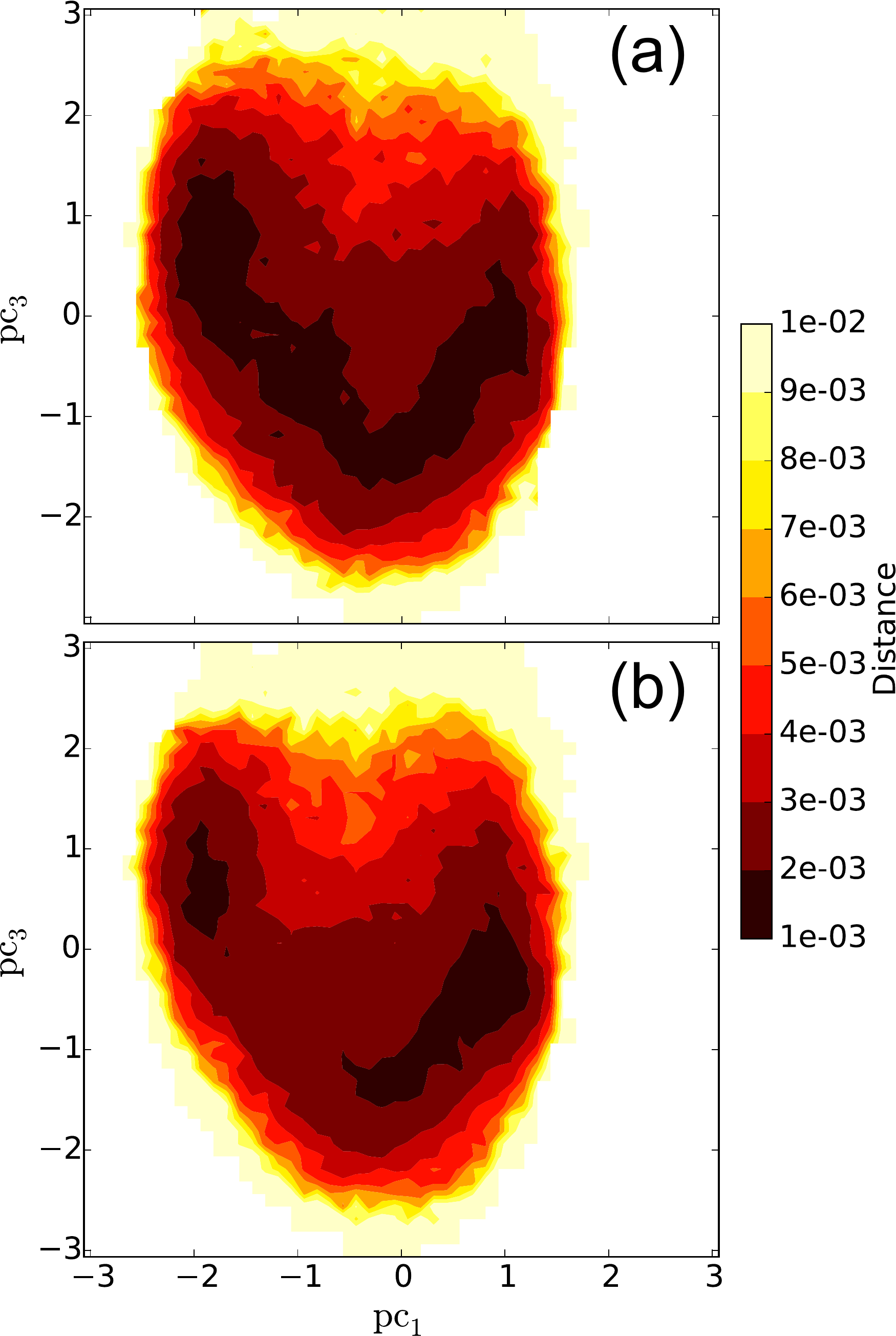}
	\caption{The value at each grid-point represents the distance, for an initial density of 1 at this grid-point, between the density $f_{8k}^\mathrm{pow}$ transferred by the $k^{th}$ power of $\hat{P_8}$ and a density $f_{8k}^\mathrm{long}$ transferred by $\hat{P}_{k\times 8}$ for (a) $k = 2$ and (b) $k = 4$.}
	\label{fig:semigroup}
\end{figure}

We further test to which extent the semigroup property (\ref{eq:semigroup}) is violated for special cases of powers $k$ of $\hat{P_\tau}$ and the corresponding matrices $\hat{P}_{k\times \tau}$.
Rather than directly calculating a distance between the matrices $(\hat{P}_\tau)^k$ and $\hat{P}_{k \times \tau}$,
we prefer to calculate, for an initial density $f_0$, the distance between the density $f_{k\tau}^\mathrm{pow} = f_0 (\hat{P_\tau})^k$ transferred by the $k^{th}$ power of $\hat{P_\tau}$ and the density $f_{k\tau}^\mathrm{long} = f_0 \hat{P}_{k \times \tau}$ transferred by $\hat{P}_{k \times \tau}$. We use the distance $d(f, g) =  \sqrt{\sum_{i = 1}^m  \pi_i (f_i - g_i)^2}$,
which is a sum of squared errors between the components of $f$ and $g$, weighted by the likelihoods $\hat{\mathbb{P}}(y_t \in B_i)$, so that errors in grid boxes less likely to be reached by the Markov process are given less weight. The distances $d(f_{k\tau}^\mathrm{pow}, f_{k\tau}^\mathrm{long})$ have been calculated for a set of $m$ initial densities associated with each grid-box in $\{B_i\}$, so that the density associated with box $B_j$ integrates to 1 in $B_j$ and to zero elsewhere. 
The distances are represented figure \ref{fig:semigroup}a and b for a lag $\tau$ of 8 days 
and for a multiplicator $k$ of $2$ and $4$, respectively (dark colors indicate small distance).

To explain the nature of the violation of the semigroup property (\ref{eq:semigroup}), let us recall that there exists only three possible candidates, namely (i) the partial observation of the dynamical system, (ii) the coarse-graining induced by the Galerkin approximation and (iii) sampling errors. Furthermore, it is shown in Appendix \ref{app:B}, that the distances plotted figure \ref{fig:semigroup} are not affected by the use of twice as more samples and that they only decrease slightly when the resolution increases (and vice versa). Thus, we can say that positive distances in figure \ref{fig:semigroup} are mostly the consequence of memory effects induced by the partial observation of the high-dimensional barotropic model.

We can observe that these memory effects are mostly important where the stationary density is small (figure~\ref{fig:histogramPC13}) and thus, where trajectories are less likely to pass by. The denser regions which are associated with meta-stability, as will be seen section \ref{sec:metastable}, seem to be less affected by memory effects. This result is in agreement with the relative constancy with lag of the leading rates, also associated with meta-stability.
Such memory effects, as well as the dependence on the lag of the spectral gap between the leading rates, should be put in perspective with stochastic modeling with Stochastic Differential Equations (SDE) and based on time-scale separation \citep{Majda1999, Majda2001, Franzke2014}. Indeed, in such models, one seeks a reduced basis for the decorrelation time of the resolved variables to be much longer than the one of the unresolved variables, in order to be able to neglect the memory effects made explicit by the Mori-Zwanzig formalism \citep{Darve2009a, Chorin2009}. If the transition matrices on the reduced phase space do not directly help in the choice of the basis on which the dynamical system can be optimally reduced, their rates can give insights on the impact of the reduction on the resolved variables.

\subsection{Meta-stable regimes as almost-invariant sets of the transfer operator}
\label{sec:metastable}

We have seen, section \ref{sec:semigroup}, that the two leading rates of figure \ref{fig:spectra} are close to zero and that a large spectral gap separates them from the rest of the rates, a configuration indicative of the presence of meta-stable regimes. This characteristic of meta-stability or persistence allows to formally define these regimes as almost-invariant sets \citep{Dellnitz1997a, Froyland2003}. We now give an extension of the definition of almost-invariant sets to sets in the reduced phase space and present an algorithm to detect them from a transition matrix $\hat{P}_\tau$.

A set $A$ of the phase space $X$ is almost-invariant if $S_\tau^{-1}(A) \approx A$, so that
\begin{eqnarray}
	\mathbb{P}(x_{t + \tau} \in A | x_t \in A) \approx 1.
\end{eqnarray}
Reformulating, the probability for a trajectory starting in a set $A$ to leave this set after a lag $\tau$ is almost-zero. These sets are thus associated with persistent or meta-stable regimes.

In the case of almost-invariant sets in the reduced phase space, we are interested in sets $E$ of $Y$, almost-invariant with
respect to the transition probabilities $\mathbb{P}(y_{t + \tau} \in E | y_t \in E)$, such that
\begin{eqnarray}
	\mathbb{P}(y_{t + \tau} \in E | y_t \in E) \approx 1.
	\label{eq:almostinv}
\end{eqnarray}
However, applying (\ref{eq:theoremA}) \cite[Theorem A]{Chekroun2014} we have that
\begin{eqnarray}
	\mathbb{P}(y_{t + \tau} \in E | y_t \in E) \approx 1 \quad \iff \quad \mathbb{P}(x_{t + \tau} \in h^{-1}(E) | x_t \in h^{-1}(E)) \approx 1.
\end{eqnarray}
This important result states that if a set $E$ is almost-invariant in the reduced space $Y$, its pre-image $h^{-1}(E)$ in $X$ is almost-invariant to the flow $S_\tau$. In other words, almost-invariant sets in the reduced phase space are images of almost-invariant, yet coarser, sets in the full phase space. Of course, these coarse-grained almost-invariant sets may not be optimal, in the sense that other, more strongly almost-invariant sets (w.r.t (\ref{eq:almostinv})) may exist but are filtered out by the observable $h$ in the same way RP resonances can be filtered out by $h$.

Based on these considerations, the transition matrix $\hat{P}_{\tau = 8}$ was used to define the meta-stable regimes objectively.
For the detection of almost-invariant sets \citep[see also,][]{Dellnitz1997a, Froyland2003, Froyland2009} we use an optimal 
Markov chain reduction \citep{Deng2011, Rosvall2008} with respect to the relative entropy rate 
\cite{Cover1991}. This type of Markov chain reduction is particularly well suited for the detection of dense almost-invariant sets (highly recurrent), since it attempts to minimize the distance between a density transferred by the reduced 
transition matrix (giving the transition probabilities between the almost-invariants, see below)
and the same density transferred by the original transition matrix.  The optimization 
was implemented using the greedy algorithm from network theory \citep{Clauset2004}, where the grid-boxes are iteratively merged to give coarser and coarser almost-invariant sets.
	
\begin{figure}
	\centering
	\includegraphics[width=8.5cm]{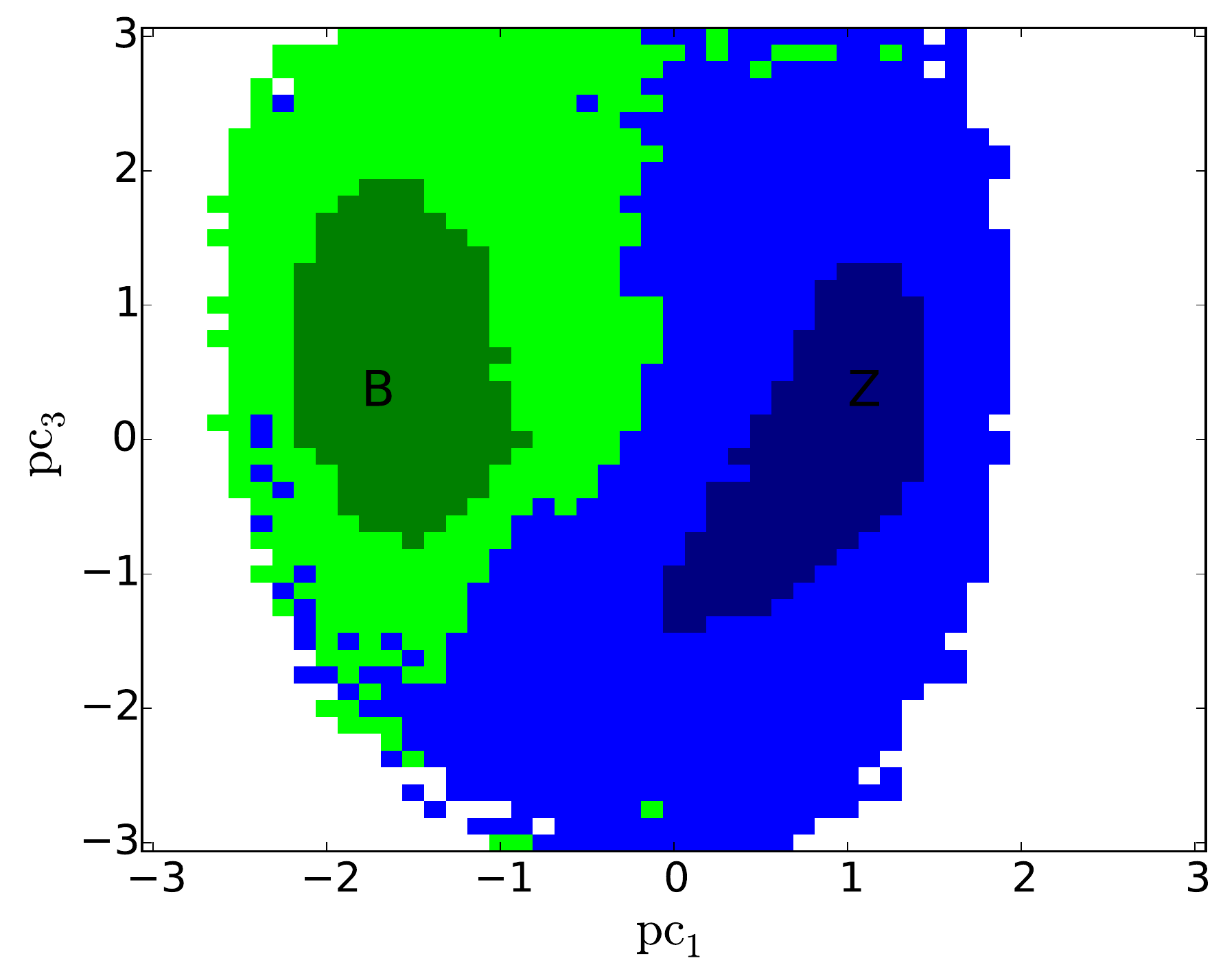}
	\caption{Two almost-invariants sets and their restrictions (in dark colors) corresponding to the blocked and the zonal regime, respectively.}
	\label{fig:regimes}
\end{figure}

In accordance with the  bimodality of the histogram (figure~\ref{fig:histogramPC13}), we have 
chosen to look for a number of almost-invariant sets $p$ of 2. These two sets are plotted in figure \ref{fig:regimes}, such that all grid boxes in green belong to the first almost-invariant set and all grid boxes in blue belong to the second one. For the family of almost-invariant sets $\{E_\beta\}$, the 2-by-2 reduced transition matrix $\hat{Q}_{\tau = 8,}$, such that $(\hat{Q}_{\tau = 8})_{\beta \gamma} = \hat{\mathbb{P}}(y_{t+\tau} \in E_\gamma | y_t \in E_\beta)$, and its stationary density $\eta$, such that $\eta_\beta =  \hat{\mathbb{P}}(y_t \in E_\beta)$, are found to be
\begin{eqnarray}
 \hat{Q}_{\tau = 8} = \begin{pmatrix}
  0.79 & 0.21 \\
  0.14 & 0.86
 \end{pmatrix},
 \quad
\eta =  \begin{pmatrix}
 0.27 \\
 0.73
 \end{pmatrix},
 \end{eqnarray}
the second almost-invariant set (in blue) being almost three times as dense as the first one (in green).

The algorithm is designed to find almost-invariant sets whose union covers the entire grid.
However, in view of the early warning problem  discussed in section \ref{sec:forecast}, we need to find 
a restriction of  the definition of the regimes $\{R_\beta\}$ to smaller regions of the grid so that the likelihood $\hat{\mathbb{P}}(y_t \in R_\beta)$ to be in any regime $R_\beta$ becomes smaller than one \cite[]{Kharin2003}.
To do so, we selected, for each almost-invariant set $E_\beta$, their grid-boxes $B_i$ maximizing the likelihood $\hat{\mathbb{P}}(y_t \in B_i, y_{t+\tau} \in E_\beta, y_{t-\tau} \in E_\beta)$ of a realization of $\mathbf{y}$ to be in $B_i$ and to come from and go to the same almost-invariant set $E_\beta$, until a sufficiently large number of boxes have been attributed to the regime $R_\beta$ to have $\hat{\mathbb{P}}(y_t \in R_\beta) = \hat{\mathbb{P}}(y_t \in E_\beta) / 2$ (until half of the almost-invariant set has been selected in terms of stationary density $\pi$). These 
restrictions are plotted in dark green and dark blue (Fig.~\ref{fig:regimes}) and define the blocked and zonal regimes, 
respectively. The probability to stay in the so-defined blocked and zonal regimes after 8 days is of $66\%$ and $70\%$,  respectively.

It is shown in the Appendices \ref{app:A}, \ref{app:B} and \ref{app:C} that these regimes are relatively robust to the limited length of the time series, to the grid resolution and to the lag, respectively.
Their definition together with the transition matrices $\{\hat{P}_\tau\}$ are used in section \ref{sec:forecast} to define an early warning indicator of a transition to the blocking regime.

\section{Preferred transition paths and early warning}
\label{sec:forecast}

\subsection{Preferred transition paths}
\label{sec:prefTrans}

Having defined the regimes, we now study the transitions between them.  Following \citet{Branstator2005}, 
we first plot the mean tendencies of the normalized principal components pc$_1$ and pc$_3$.  The tendencies 
were calculated for each principal component using a  
finite difference scheme such that $\Delta \text{pc}_i(t) = (\text{pc}_i(t + \Delta t) - \text{pc}_i(t)) / \Delta t$, where $\Delta \text{pc}_i$ is the approximate tendency of the $i^{th}$ principal component and 
$\Delta t$ is the time step. An estimate of the mean 
tendency for each grid box was then calculated by averaging over all the realizations of $\mathbf{y}$
in this grid box. 

\begin{figure}
	\centering
	\includegraphics[width=8.5cm]{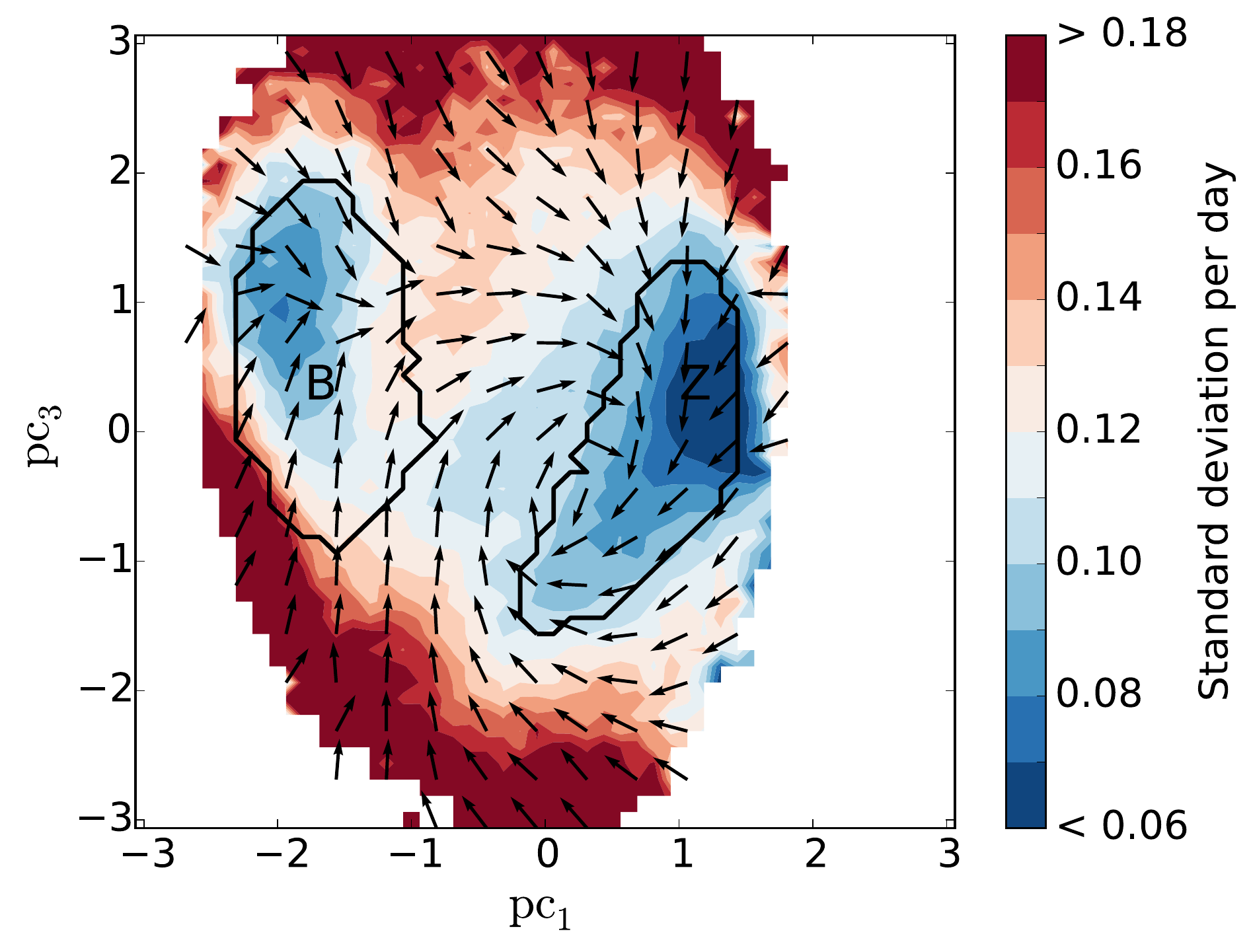}
	\caption{Mean tendency of the normalized principal components calculated using centered 
	differences for $\Delta t = 8$ days.  Arrows represent the direction 
	and the colors represent the magnitude (in standard deviation per day).
	As for the following figures, the black contours delimitate the blocked regime, marked by the letter B,
	and the zonal regime, marked by the letter Z.}
	\label{fig:ten8}
\end{figure}

The mean tendency for a time-step $\Delta t$ of 8 days is plotted figure \ref{fig:ten8}.
It can be seen as a composition of a clockwise rotation and two sinks. This result corroborates 
both the meta-stability of the regimes and the existence of preferred transition paths between them, reminiscent 
of a pseudo-periodic orbit   \citep{Plaut1994, Crommelin2003}. Indeed, the rotation is such 
that typical trajectories leaving the zonal regime (blocked regime) to go to the blocked regime (zonal regime)
transit through negative values (positive values) of pc$_3$. Furtermore, the correspondence 
between the sinks with low-values of mean tendency and the regimes is striking, in particular for the zonal 
regime. We have seen in section \ref{sec:semigroup} that the memory effects are relatively weak in the 
region of the regimes. In the limit when these effects can be neglected, the reduced dynamics
can be modeled by an SDE and the mean tendency gives an approximation of the drift term
involved in the Fokker-Planck equation \citep[together with diffusion, not calculated here,][]{Berner2005a, Gardiner2009} which generates the 
semigroup of transfer operators associated with the SDE. The correspondence between weak tendency and almost-invariance is thus not coincidental.

To further support the existence of preferred transition paths from one regime to the other,
we have calculated, for each grid box, the likelihood $\hat{P}_{ZB}$ ($\hat{P}_{BZ}$)
that a trajectory starting in the zonal regime (blocked regime) and passing through this grid box
reaches the blocked regime (zonal regime) before the zonal regime (blocked regime). 

\begin{figure}
	\centering
	\includegraphics[width=8.5cm]{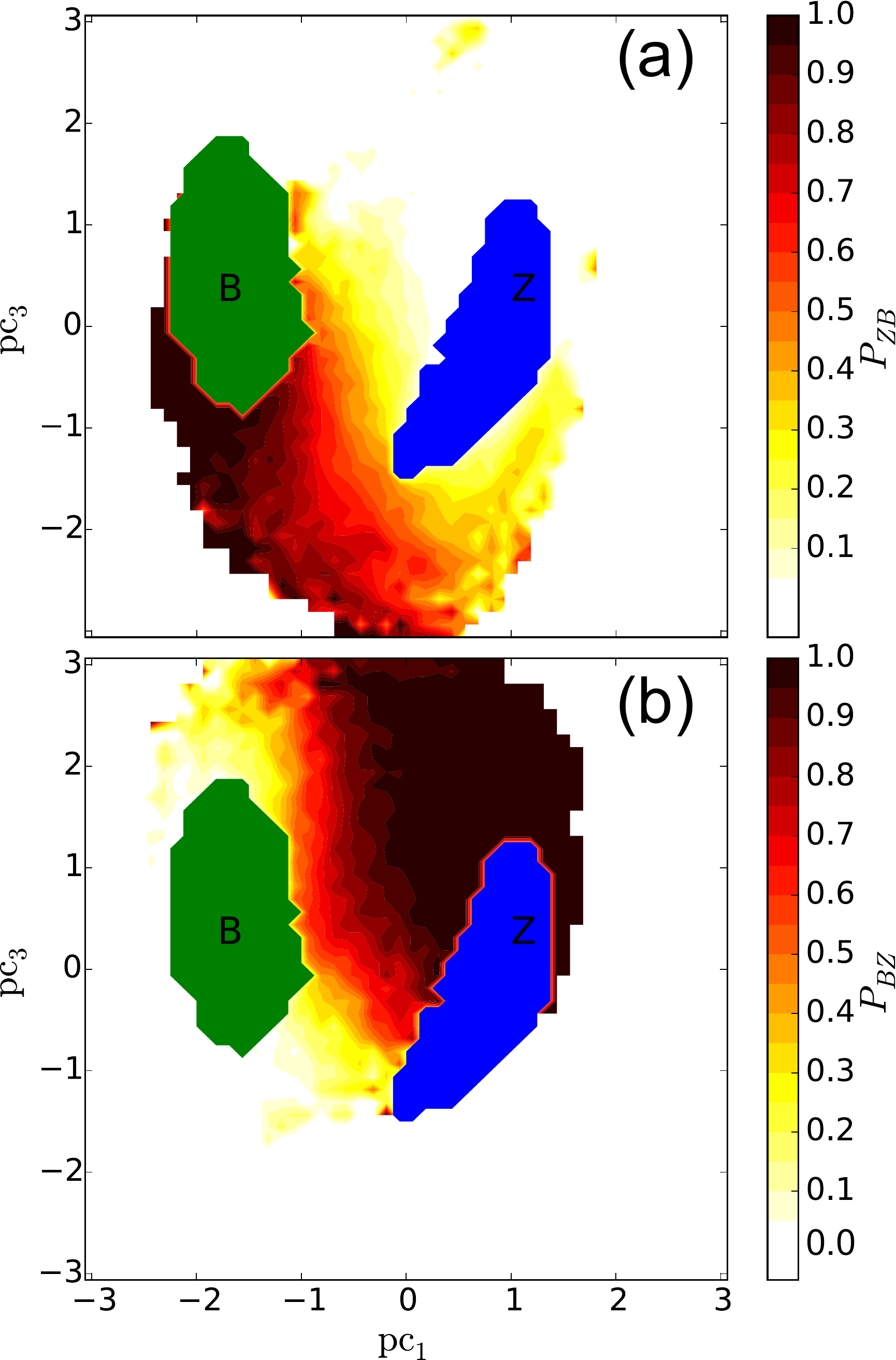}
 	\caption{Likelihood, for each grid box, to reach (a) the blocked regime before the zonal regime
	and (b) the zonal regime before the blocked regime.}
	\label{fig:probB}
\end{figure}

The resulting likelihoods are plotted in figure \ref{fig:probB}.  In agreement with the tendency, the trajectories going from the zonal to the blocked regime are more likely to do so through low values of the $3^{rd}$ principal component while trajectories going from the blocked to the zonal regime favor high values of the $3^{rd}$ principal component.

\subsection{Early warning indicator}
\label{sec:early}

The presence of preferred transition paths from the zonal to the blocked regime, as well as
the weak mixing associated with the existence of rates close to zero (cf. section \ref{sec:spectra}), suggests that there is  potential skill in 
predictability of transitions to the blocked regime. Note that more than trying to predict when a trajectory
will leave the zonal regime, we want to use the fact that trajectories 
leaving the zonal regime are more likely to go to the blocked regime if they transit through negative 
values of pc$_3$. We therefore use the transition matrices for different lags to 
provide an approximation of the transfer of densities in the reduced phase space by the flow
and to build an early-warning indicator of transitions to the blocked regime.

The quality of the indicator will depend on (i) the predictability of the full system (in terms of sensitive dependence to initial conditions or rate of decay of correlations) and (ii) the validity of the Markov approximation for the reduced dynamics.
While the former is determined by the dynamical system alone, the latter depends on the choice of the observable.
We have seen in section \ref{sec:spectraMetastable} that the violation of the semigroup property indicates that the reduced dynamics are overall non-Markovian.
However, the constancy of the leading rates also indicates that motions associated with the meta-stable regimes
behave as Markovian, suggesting that densities transferred by the estimated transition matrices
could be used to give a probability of reaching one of the  regimes.

The early-warning indicator is thus designed as such. Whenever the observed trajectory (in our case, the simulation) leaves the zonal regime,
we define an initial density $f_{B_i, \tau = 0}$,  integrating to 1 over grid-box $B_i$ to which the last observation belongs and to zero elsewhere (one could
instead define an initial density spreading over several boxes to account for uncertainties in the observation).
Next, the Markov approximation $f_{B_i, \tau = 1}$ of the transfer of $f_{B_i, \tau = 0}$ by the flow after a lag $\tau$ of 1 day
is calculated as $f_{B_i, \tau = 0} \hat{P}_{\tau = 1}$.
We then calculate the likelihood
$\hat{\mathbb{P}}(y_{\tau = 1} \in R_\mathrm{block} | y_0 \in B_i)
= \sum_{B_j \in R_\mathrm{block}} (f_{B_i, \tau = 1})_j$
for a realization $y_{t + 1}$ to belong to the blocking regime knowing that $y_t$ belongs to grid-box $B_i$.
If this estimated probability exceeds a given critical probability $p_c$, an alarm of transition to the blocked regime after a lag $\tau_\mathrm{alarm}$ of 1 day is given. Otherwise, the same process is repeated for $\tau = 2, 3, ...$ until an alarm is given or a limit lag $\tau_\mathrm{max}$ is reached, after which we wait for the next observation to run the forecasting system.

\begin{figure}
	\centering
	\includegraphics[width=17cm]{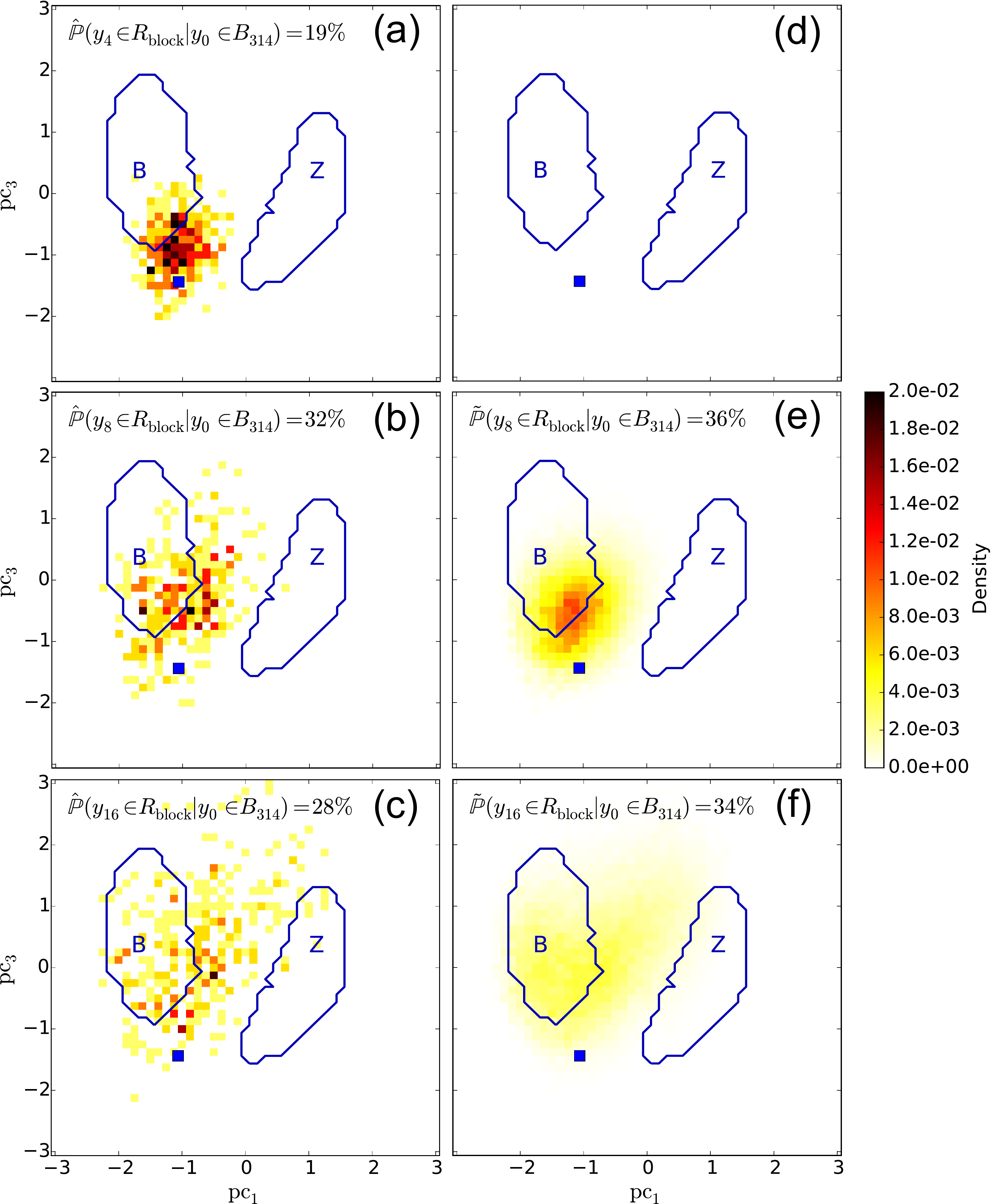}
	\caption{Markov approximation $f_{B_{i = 314}, \tau}$ of the transfer of the initial density $f_{B_{i = 314}, \tau = 0}$
	for a lag $\tau$ of (a) 4 days, (b) 8 days and (c) 16 days.
	The initial density is plotted in panel (d), integrating to one over box $B_{i = 314}$ and to zero elsewhere.
	The densities
	transferred using the square of $\hat{P}_{\tau = 4}$ and of $\hat{P}_{\tau = 8}$ are plotted panel (e) and (f), respectively.
	The corresponding likelihoods to reach the blocked regime are written in the top left of each panel.}
	\label{fig:transDen}
\end{figure}

To illustrate this process we show in figure \ref{fig:transDen} the transferred density for 
grid box 314 marked as a 
blue square and lying in the region of preferred transition paths between the zonal and the blocked regimes.
This initial density $f_{B_{i = 314}, \tau = 0}$, is plotted alone panel d).
The densities $f_{B_{i = 314}, \tau = 4}$, $f_{B_{i = 314}, \tau = 8}$ and $f_{B_{i = 314}, \tau = 16}$
transferred using the transition matrices $\hat{P}_{\tau = 4}$, $\hat{P}_{\tau = 8}$ and $\hat{P}_{\tau = 16}$ are plotted
panel (a), (b) and (c), respectively.
The densities $f_{B_{i = 314}, \tau = 2 \times 4}$ and $f_{B_{i = 314}, \tau = 2 \times 8}$ transferred
using the square of the transition matrices $\hat{P}_{\tau = 4}$ and $\hat{P}_{\tau = 8}$ are also plotted
panel e) and f), respectively, to show how the semigroup property can be violated.
The corresponding likelihoods to reach the blocked regime are written in the top left of each panel.

\begin{figure}
	\centering
	\includegraphics[width=8.5cm]{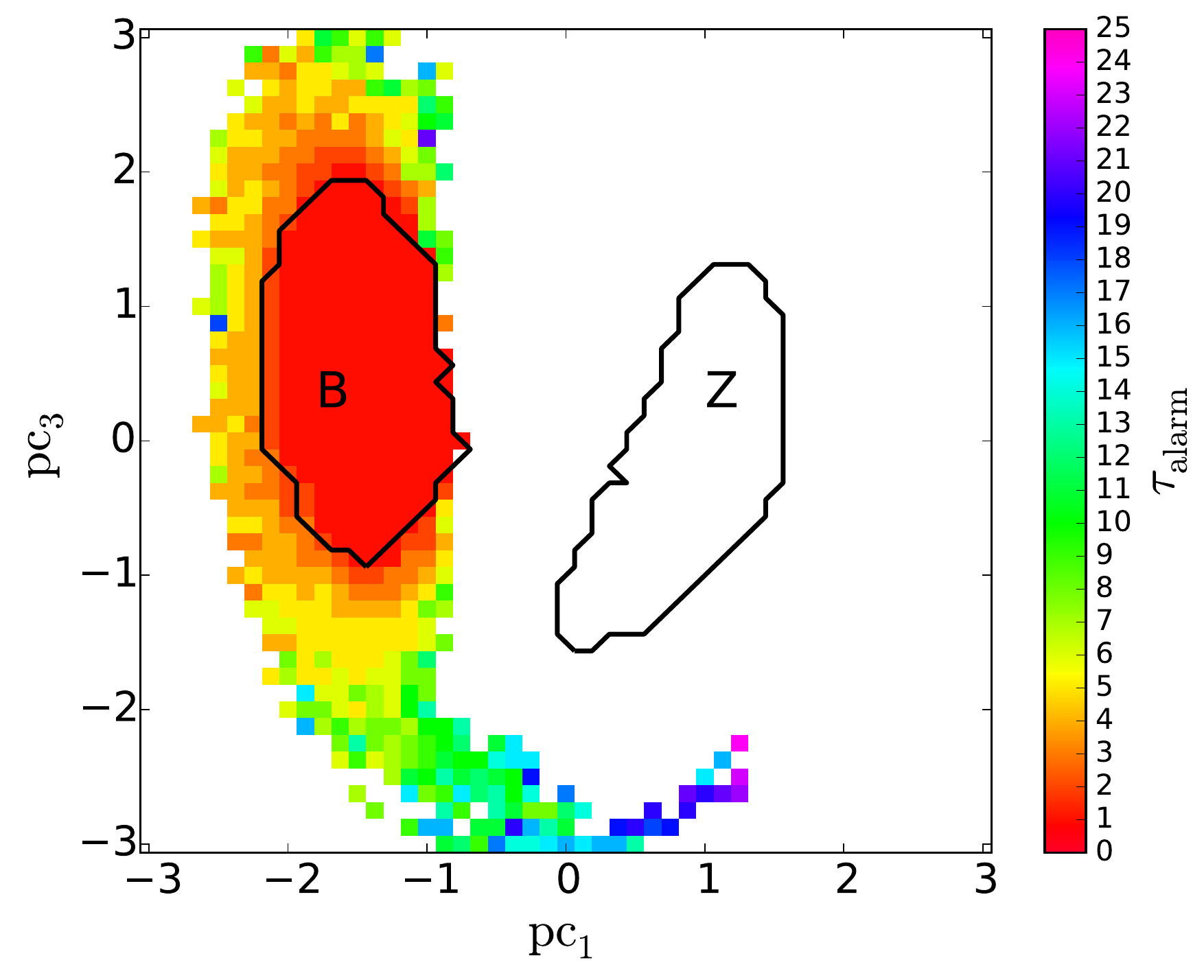}
	\caption{Predicted time of transition to the blocked state $\tau_\mathrm{alarm}$, in days,
	for each grid box where an alarm is given for a critical probability $p_c$ of 0.3.}
	\label{fig:tauMap}
\end{figure}

We can see that if a critical probability $p_c$ of 0.3 would be chosen, an alarm of transition to the blocking regime would be
raised for $\tau = 8$ days (figure~\ref{fig:transDen}b), as described in the previous paragraph.
Figure \ref{fig:tauMap} shows the grid boxes for which an alarm would 
be given if a critical probability $p_c$ of 0.3 was used and the respective lag $\tau_\mathrm{alarm}$ after which the 
transition is predicted to occur. In this case, alarms are mainly flagged for trajectories passing through 
grid boxes close to the blocked regime or in the region of low principal component pc$_3$.

The quality of the early warning indicator of transition to the blocked regime was tested over all trajectories starting 
when leaving the zonal regime and ending when reaching any of the regimes.
In order to take into account sampling errors in the estimation of the transition matrices,
the test was performed on a second $500,000$-day-long simulation which was ran
taking as initial state the last state of the original simulation from which the transition matrices were estimated.
Out of the $11759$ trajectories starting from the zonal regime, $1993$ 
reached the blocked regime, which we call an \emph{occurrence} event (O), while $9766$ reached the zonal regime, 
which we call a \emph{non-occurrence} event ($\overline{\text{O}}$).

When testing a forecast system, one is interested in how many alarms (A) were given when the event actually occurred, the number of so-called \emph{hits}, and how many did not occur, the number of so-called \emph{false alarms}. In the case of a hit, the event occurs at a given time 
$\tau_{\mathrm{occ}}$ after the alarm is given, so that not only the occurrence of the event 
should be forecasted, but also when it will occur. For easy reference, the different 
cases are shown in Table~\ref{tab:events}.
For example, a forecast that a blocking event will occur in a month although it will truly occur only a week later is not of much help. For the purpose of assessing the skill of a forecast not only in terms of occurrence but also in terms of precision of the forecasted date of occurrence, we need to define a tolerance $\epsilon$, in days, such that a hit (H) is granted only when an alarm is given with a prediction time $\tau_\mathrm{alarm}$ such that $\tau_{\mathrm{occ}} - \epsilon \le \tau_{\mathrm{alarm}} \le \tau_{\mathrm{occ}} + \epsilon$. If, however,  the event is predicted to happen too early ($\tau_{\mathrm{alarm}} < \tau_{\mathrm{occ}} - \epsilon$), it is counted as a \emph{false alarm of type II} (FA2), while if the event is predicted to happen too late ($\tau_{\mathrm{alarm}} > \tau_{\mathrm{occ}} + \epsilon$), it is counted as a \emph{missed alarm of type II} (MA2). When the event did not occur and no alarm ($\overline{\text{A}}$) was given, we count a \emph{correct rejection} (CR).  When an event occurs but no alarm is given, we count a \emph{missed alarm of type I} (MA1) and when an event is predicted but does not occur, we count a \emph{false alarm of type I} (FA1).  
\begin{table}
	\caption{Overview of the different cases which can occur depending on the alarm given and the event occurring.}
	\begin{tabular}{| l | c | c | c | c |}
		\hline
		& A & A & A & $\overline{\text{A}}$ \\
		\hline
		& $\tau_{f} < \tau_{o} - \epsilon$ & $|\tau_{f} - \tau_{o}| \le \epsilon$ & $\tau_{f} > \tau_{o} + \epsilon$ & \\ \hline
		O & FA2 & H & MA2 & MA1 \\ \hline
		$\overline{\text{O}}$ & FA1 & FA1 & FA1 & CR \\ \hline
	\end{tabular}
	\label{tab:events}
\end{table}

To assess the skill of the  forecasts based on the categories defined in  Table \ref{tab:events}, we 
adapted the original Peirce skill score (PSS) \citep{Peirce1984, Stephenson2000}  to forecasts  
including the time of  occurrence. In our case, we define the PSS as
\begin{eqnarray*}
	S_{Peirce}(p_c) = \text{HR}(p_c) - \text{FAR}_1(p_c),
\end{eqnarray*}
where $\text{HR}(p_c) = \text{H}(p_c) / \text{O}(p_c)$ is the hit rate, defined as the ratio of the 
number of hits over the number of occurrences,  and $\text{FAR}_1(p_c) = \text{FA1}(p_c) / \overline{\text{O}}(p_c)$ 
is the  false alarm rate of type I. The hit rate gives the likelihood of giving an alarm when the event occurs, while 
the false alarm rate of type I gives the likelihood that an alarm is given but the event does not occur.  If the hit rate 
exceeds the false alarm rate the  PSS is positive and the forecast has skill. A PSS of 1 is reached for a perfect 
forecast, when alarms are only raised when an event will actually occur and with the right predicted time of occurrence. 

\begin{figure}
	\centering
	\includegraphics[width=8.5cm]{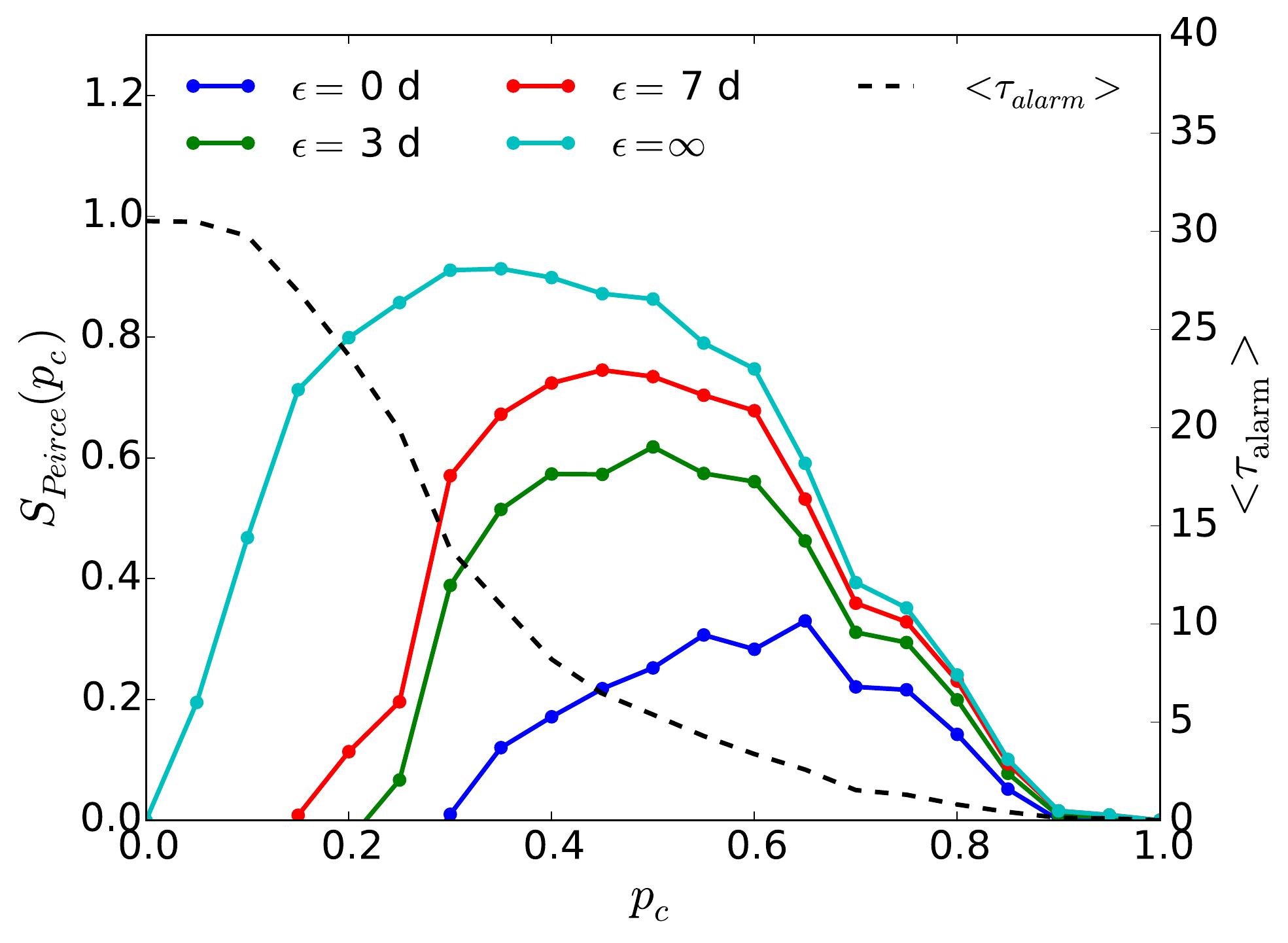}
	\caption{Peirce skill score of the probability forecast of reaching the blocked regime 
	versus the critical probability $p_c$  for different tolerances $\epsilon$ (straight lines).
	The average time lag in days for which an alarm is given is also plotted in black dashed line.}
	\label{fig:Peirce}
\end{figure}

The PSS is plotted figure \ref{fig:Peirce} versus the critical probability $p_c$ and for 
different tolerances $\epsilon$. A tolerance of $\infty$ means that the precision of the predicted date of occurrence is
not taken into account in the skill score. It is evident that as the tolerance increases, the skill score increases as well.
The PSS reaches a maximum for a critical probability around 0.5, depending 
on the tolerance.
Such scores have to be put in perspective with the predicted time of occurrence $\tau_\mathrm{alarm}$.
Indeed, a successful prediction of occurrence of a blocking event $1$ day ahead is counted as a hit,
but is not of much practical use.
For this reason, we have also represented the average prediction time $\tau_\mathrm{alarm}$ as a black
dashed line on the same figure \ref{fig:Peirce}.
It is of around two weeks for a critical probability of 0.3 and around one week for a critical probability of 0.45.
Thus, depending on the final purpose of the forecast system, a compromise has to be found between
the skill of the forecast system and how many days ahead an event is predicted on average.
Choosing a critical probability of 0.3 would give a reasonably good skill score of 0.4 (better than a no-skill forecast)
with an average prediction time of 2 weeks.

To summarize, the main advantage of a forecast system relying on the transfer of densities 
is that it constitutes a very cheap way to account for the sensitive dependence on initial 
conditions of a chaotic dynamical system. 

\section{Energetics of the transitions}

In this section we focus on the remaining question on how the dynamics of the barotropic model can 
explain (i) the persistence of each regime and (ii) the preferred transition paths from the zonal to the blocked 
regime through high values or $\textnormal{pc}_3$. 

To help clarifying these issues the hemispheric energy budget of the model is studied. First, the 
fields  are decomposed in a $\bar{\tau} = 8$ days running mean and a deviation  from it.  This 
decomposition yields for the streamfunction 
\begin{eqnarray}
	\begin{split}
		\psi & = \overline{\psi} + \psi'  \quad \text{with} \quad \overline{\psi} = 
		\frac{1}{\bar{\tau}} \int_{t-\bar{\tau}/2}^{t+\bar{\tau}/2} \psi ~  d t'.
	\end{split}
	\label{eq:decomp}
\end{eqnarray}
Inserting (\ref{eq:decomp}) into equation (\ref{eq:bve})and applying the running mean 
gives the equation of the mean relative vorticity
\begin{eqnarray}
	\frac{\partial{\nabla^2\overline{\psi}}}{\partial{t}}
	 + \mathcal{J}(\overline{\psi}, \nabla^2 \overline{\psi} + f + h)
	+ \overline{\mathcal{J}(\psi', \nabla^2\psi')}
	 = -k_1 \nabla^2 \overline{\psi} + k_2 \nabla^8 \overline{\psi} + \nabla^2 \psi^*.
	\label{eq:modelMean}
\end{eqnarray}
Subtracting (\ref{eq:modelMean}) from (\ref{eq:bve}) gives the equation of the deviation 
from the running mean as 
\begin{eqnarray}
		\frac{\partial{\nabla^2\psi'}}{\partial{t}}
		 + \mathcal{J}(\overline{\psi}, \nabla^2\psi') + \mathcal{J}(\psi', \nabla^2 \overline{\psi} + f + h) \nonumber \\
		 + \mathcal{J}(\psi', \nabla^2\psi') - \overline{\mathcal{J}(\psi', \nabla^2\psi')} 
		 = -k_1 \nabla^2 \psi' + k_2 \nabla^8 \psi'.
	\label{eq:modelPerturb}
\end{eqnarray}

In order to obtain the equation of the hemispheric average of the mean kinetic energy $\overline{E}=  
<\frac{\overline{u}^2 + \overline{v}^2}{2} >$, with  $< \cdot >$ denoting the hemispheric average 
$\frac{1}{2 \pi} \int_0^{2 \pi} \int_{0}^{\pi/2} \cdot  \cos \phi \textnormal{d}\phi\textnormal {d}\lambda$,  
equation (\ref{eq:modelMean}) is multiplied by $\overline{\psi}$ and averaged hemispherically, 
giving
\begin{eqnarray}
	\frac{\partial \overline{E}}{\partial{t}} 
	 = <\overline{\psi} \text{ } \overline{\mathcal{J}(\psi', \nabla^2 \psi')}>
	 - 2 k_1 \overline{E} - k_2 <\overline{\psi} \nabla^8 \overline{\psi}>
	 - <\overline{\psi} \nabla^2 \psi*>  .
	\label{eq:MKE}
\end{eqnarray}
The first term on the right hand side is equal to the opposite of the sum of the 
Reynolds' stress terms which represent a conversion of mean to eddy kinetic energy.
Finally, multiplying (\ref{eq:modelPerturb}) by $\psi'$, applying the running mean 
and averaging over the hemisphere gives the equation of the global eddy 
kinetic energy $E' =  <\frac{\overline{u'^2} + \overline{v'^2}}{2} >$
\begin{eqnarray}
	\frac{\partial E'}{\partial t} 
	 = - <\overline{\psi} \overline{\mathcal{J}(\psi', \nabla^2 \psi')}>
	 - 2 k_1 E' - k_2 <\psi' \nabla^8 \psi'>.
	\label{eq:EKE}
\end{eqnarray}
The terms  in the equations (\ref{eq:MKE}) and (\ref{eq:EKE}) were calculated from the model 
simulation results and we could verify that the calculated tendencies equated the sum of the right 
hand side terms but for a small error of up to  $13\%$ of the standard deviation of the tendencies 
due  to the running average of a deviation not being exactly zero. 

\begin{figure*}[t]
	\centering
	\includegraphics[width=17cm]{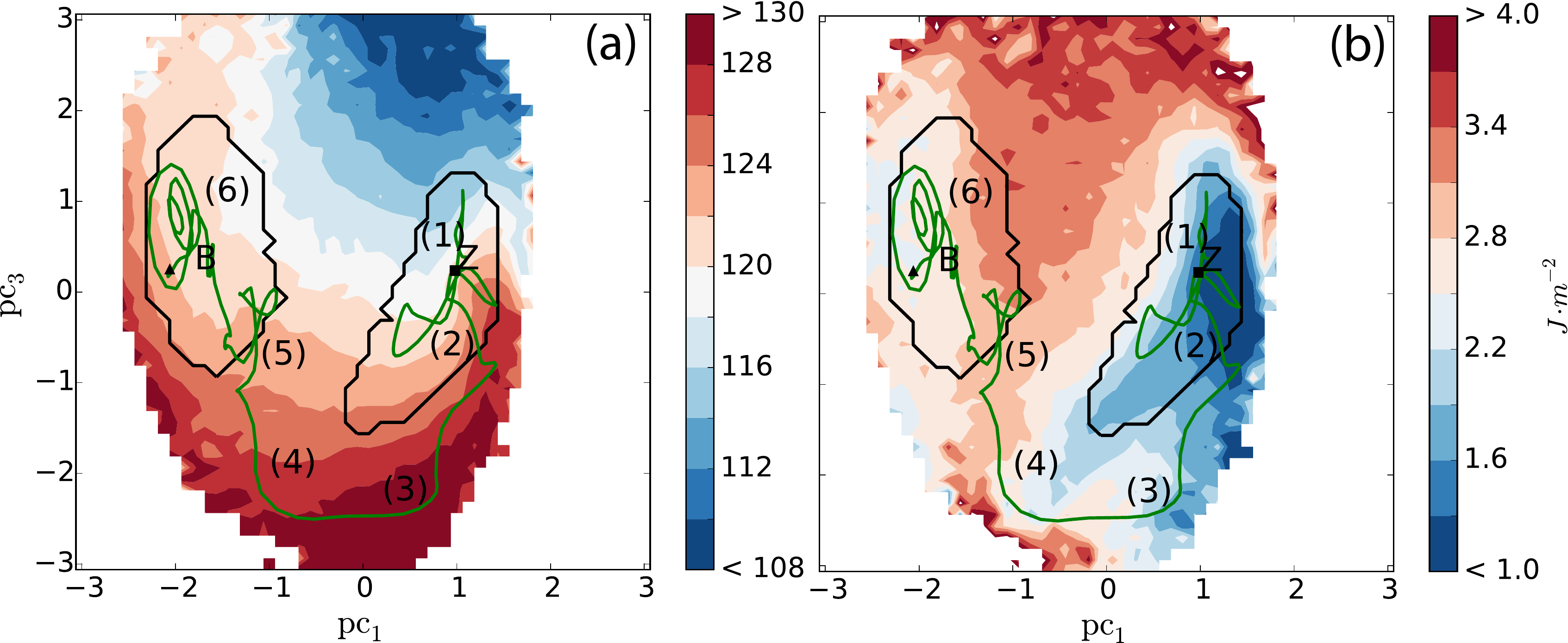}
	\caption{(a) Hemispheric mean kinetic energy $\overline{E}$ in Jm$^{-2}$, 
	(b) Hemispheric eddy kinetic energy $E'$ in Jm$^{-2}$.
	The green line, starting with a black square and ending with a 
black triangle, represents a 200 days long trajectory transiting from 
the zonal to the blocked regime.}
	\label{fig:projE}
\end{figure*}

\begin{figure*}[t]
        	\includegraphics[width=17cm]{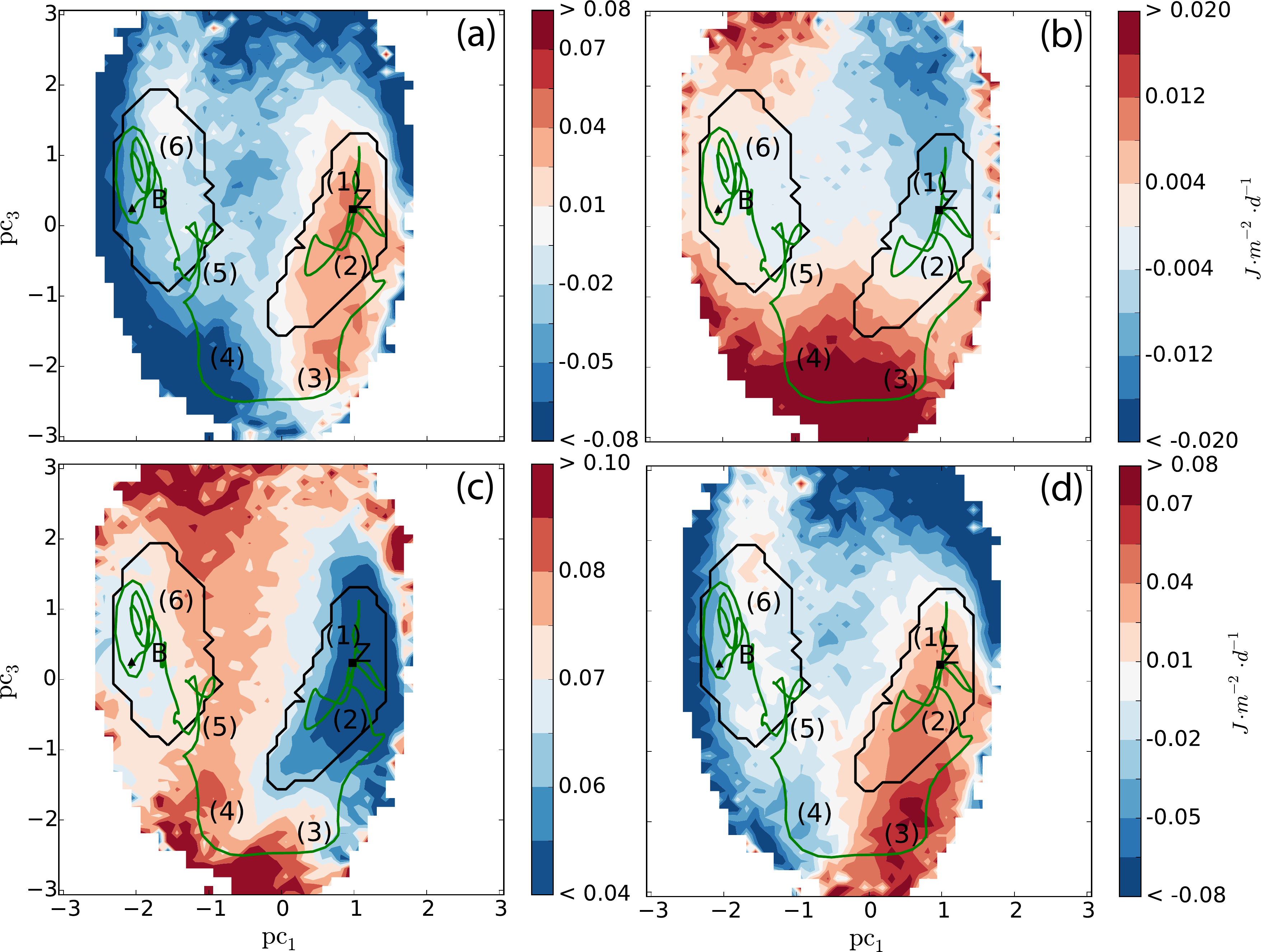}
	\caption{(a) Tendency of the hemispheric mean kinetic energy in Jm$^{-2}$day$^{-1}$,
	(b) Tendency of the hemispheric eddy kinetic energy in Jm$^{-2}$day$^{-1}$,
	(c) Hemispheric Reynold's stresses (conversion from mean to eddy kinetic energy when 
	positive) in Jm$^{-2}$day$^{-1}$,
	(d) Sum of hemispheric forcing and dissipation in Jm$^{-2}$day$^{-1}$.
	The green line, starting with a black square and ending with a 
black triangle, represents a 200 days long trajectory transiting from 
the zonal to the blocked regime.}
	\label{fig:projBudget}
\end{figure*}

The energetics of the transitions can be studied by plotting the kinetic energies (Fig.~\ref{fig:projE}) 
and the terms in the energy  budget (Fig.~\ref{fig:projBudget}) averaged for each grid box of the 
reduced phase space. To these plots, a 200 days long trajectory transiting smoothly from 
the zonal to the blocked regime is added in green, starting with a black square and ending with a 
black triangle. It is first interesting to notice that low values of $E'$  coincide rather well with our definition 
of the regimes (Fig.~\ref{fig:projE}b). That fact that the eddies are weak in the neighborhood of the regimes is 
mostly explained by low values of conversion to eddy kinetic energy (Fig.~\ref{fig:projBudget}c), in 
particular for the zonal regime, and additionally by a negative forcing for the blocked regime 
(Fig.~\ref{fig:projBudget}c-d).  This stabilization of the flow in the region of the regimes is a good physical
candidate to explain their persistence. 


We have seen (section \ref{sec:forecast}), that typical trajectories from the zonal to the blocked 
regime transit through the region of negative  $\textnormal{pc}_3$. Figures \ref{fig:projE} and 
\ref{fig:projBudget}  allow us to give a typical scenario for such a transition. Each step of the 
following scenario is marked in figures \ref{fig:projE} and \ref{fig:projBudget} by the corresponding 
number. 
\begin{enumerate}
	\item Starting in the zonal regime, the Reynolds' stress terms are small and  
	the mean flow is stable. However, the positive forcing induces an increase of the 
	mean kinetic energy $\overline{E}$.
	\item As $\overline{E}$ increases and the forcing persists, the trajectory evolves to lower 
	value of $\textnormal{pc}_3$ and eventually leaves the zonal regime.
	\item The trajectory then reaches a region of $\textnormal{pc}_1$ close to zero and low $\textnormal{pc}_3$. The forcing is still strong but the Reynolds stress terms begin to increase because the strongly sheared flow becomes barotropically unstable for lower values of $\textnormal{pc}_3$, so that the total eddy kinetic energy $E'$ continues to increase. 
	\item As the trajectory reaches lower values of $\textnormal{pc}_1$, the forcing reverses but the Reynolds' stress terms continue to increase as the barotropic eddies which emanated at the previous step develop, so that $\overline{E}$ is 
	converted to $E'$ which continues to increase.
	\label{step:stabilize1}
	\item The trajectory then goes to larger values of $\textnormal{pc}_3$ where the forcing is negative and energy is removed both from the mean flow and from the eddies so that the barotropic eddies decay.
	\label{step:stabilize2}
	\item The trajectory eventually reaches the blocked regime for lower values of $\textnormal{pc}_1$ where $E'$ 
	decreases due to the negative forcing coincident with relatively small Reynolds' stress terms. The mean flow is 
	once again relatively stable, although not as much as for the zonal regime, as can be seen for the relatively large Reynolds stress terms in the region of largest $\textnormal{pc}_1$ inside the blocked regime.
\end{enumerate}

This scenario is consistent with the mechanisms of chaotic itinerancy \citep{Itoh1996}, heteroclinic 
connections \citep{Weeks1997, Crommelin2003} and almost-invariant sets bounded by invariant 
manifolds \citep{Froyland2009}. Indeed, when the reduced state is in the zonal regime, the flow is 
relatively stable as it belongs to what would be the basin of attraction of the zonal regime. However, 
as it moves towards the neighborhood of positive forcing, energy is given to the mean flow, the 
horizontal shear increases and the flow becomes unstable to small perturbations. The 
increasing Reynolds' stress term indicates  that the perturbation grows,  starts to interact 
with the mean flow and enables the   state to  leave the basin of attraction of the 
zonal regime.

\section{Summary, Discussion and Conclusions}

Using concepts of transfer operators of dissipative dynamical systems, stochastic reduction
and meta-stable regimes, we developed  an early 
warning indicator of midlatitude atmospheric regime transitions. It was applied 
here to  the transitions  between a zonal and blocked flow in a barotropic hemispheric 
atmosphere model. 

Transfer operators yield the evolution of densities induced by the flow.
Markov operators have been estimated on the reduced phase space to
approximate the point spectrum of the generator of the transfer operators, namely the 
Ruelle-Pollicott resonances. Their approximation for different lags have been used to assert the
presence of slow dynamics associated with meta-stable regimes and the impact of memory
effects on these motions. The presence of rates close to zero
and separated from the rest of the spectrum supports the existence of almost-invariant
sets associated with meta-stable regimes and with time scales of three to six weeks.
Furthermore, the relative constancy of these rates showed that  memory effects are  
weak in the meta-stable regimes and along the transition path from the zonal to the blocked regime.

In order to objectively define these regimes, we have developed an algorithm
for the computation of almost-invariant sets based on  earlier work in \citep{Dellnitz1999a, 
Froyland2003, Froyland2009}, as well as  on optimal Markov chain reduction 
\citep{Deng2011, Rosvall2008} and greedy  optimization in networks \citep{Clauset2004}. 
The algorithm attempts to minimize a measure of the distance between the reduced Markov 
operator (giving the transition probabilities between the almost-invariants) and the original one
and allows the detection of sets which are both recurrent and persistent. The algorithm
has enabled us to robustly define the blocked and the zonal regimes.

Compared to spectral almost-invariant detection algorithms such as developed earlier 
by \citep{Dellnitz1997a, Dellnitz1999a, Froyland2003, Froyland2009}, which are based on the decomposition
of the leading eigenvectors of the transfer operator into characteristic functions supporting the
almost-invariant sets, our algorithm is expected to perform better in the detection of more than two almost-invariant sets,
since the latter uses an aggregative implementation rather than a divisive one like the spectral algorithms. 
For example, a similar algorithm \citep{Rosvall2008} has been used for the detection of a dozen of communities, 
associated with spatial patterns of variability, in a correlation network of sea surface 
temperature \citep{Tantet2014}. In this study, we were only interested in the bi-partition 
problem and our algorithm showed comparable performance in terms of invariance with 
respect to that in \citep{Froyland2009}.

The transfer operator based algorithms are also comparable to Hidden Markov Models 
\citep[HMM,][]{Majda2006, Franzke2008}, with the difference that the former are non-parametric 
(no assumption is made on the distribution of the reduced states). Contrary to HMM, our algorithm 
detects hard clusters (non-overlapping sets) and takes as input the transfer operator 
rather than directly the time series of the observable. In theory, it would be possible to adapt 
the algorithm to soft clustering (to allow overlapping between the clusters) but the 
optimization problem would become harder as it would necessitate another algorithm 
than the (combinatorial) greedy algorithm.

The energy budget of the model showed that striking similarities exist between regions
of low EKE and regions of almost-invariance, weak tendency and small memory effects.
Low EKE is indicative of stability of the mean flow and thus
of almost-invariance, low tendency as well as weaker memory effects due to the small amplitude of the faster
unresolved variables.

The early warning indicator is based on a forecasting scheme involving the evolution of densities
by the estimated transfer operator in a two-dimensional 
reduced phase space spanned by two EOFs of the model. It relies on the
Markov approximation of the evolution of an initial density in the neighborhood of the latest observation
of the system and on the estimation of the probability to reach the blocked regime.
A warning is broadcasted for a lag $\tau_\mathrm{alarm}$ if this probability exceeds a prescribed  critical probability.
The quality of the early warning indicator, as 
measured by the  Peirce Skill  Score, is highest for a critical probability of about 0.5 but
a smaller critical probability of 0.3 allows to emit warnings two weeks ahead of the event, on average.
A next step is to investigate  how this promising indicator would perform
for more realistic models of atmospheric flow transitions. 

While the model here has obvious deficiencies (e.g. the lack of the representation 
of baroclinic instability), it is one of the midlatitude atmospheric models for which
regime transitions are found and hence forms a nice test model for the development 
of the early warning indicator. The computational procedure can in principle be carried
out with a General Circulation Models (GCM) exhibiting regime behavior 
\citep{Dawson2012, Dawson2014}, if  sufficiently long simulations can be performed. 
Note that in our case, we used a 500,000-day-long  
simulation ($\sim 1300$ years) to assure the significance of our statistics. However,
the dominant part of the spectrum was found to be robust to the use of only $50 \times 365$
samples which is the typical size of an operational GCM or a reanalysis record.
The question remains if the transfer operator based early warning indicator would
perform well with nearly Gaussian GCM or observational data \citep{Dawson2012, Dawson2014, 
Majda2006}, where it is not clear whether preferred transition paths between the 
meta-stable regimes exist. 

Another application could be to estimate transfer operators from a long run of an operational weather prediction GCM
and use these operators in parallel with a deterministic run of the GCM, in order to account 
for both uncertainty in the deterministic forecast  \citep{Palmer2005}, and to
give an early warning indicator of transition to a new meta-stable regime.

Finally, the early warning indicator  presented here certainly extends those based on 
critical slowdown \cite[]{Scheffer2009},  such  as the increase in variance and lag-1 
autocorrelation, which definitely  have problems in high-dimensional phase space. 
Even the network based indicators  \citep{VanderMheen2013, Viebahn2014a, Feng2014, 
Tirabassi2014a}  are not readily extended   to high-dimensional systems.  It is therefore 
hoped that the techniques in this paper  will find  application in  many types of chaotic 
high-dimensional systems as found in physics and  engineering.

\begin{acknowledgements}
The authors would like to thank Daan Crommelin for sharing his version of the model and
giving us valuable comments on the manuscript, in particular for section \ref{sec:spectra}.
We also deeply thank Mickael Chekroun for explaining the essence of the article \citep{Chekroun2014},
his comments on the robustness of the results and for the inspiring discussions regarding the ergodic theory of
dissipative systems.
Finally, the authors are very thankful to the reviewers whose careful comments greatly contributed to the
improvement of the article, in particular regarding section \ref{sec:spectraMetastable}. The quality of the 
present paper would not have been the same without the advises of these people.  The authors would like to 
acknowledge the support of the LINC project (no. 289447)  funded by EC's Marie-Curie ITN 
program (FP7-PEOPLE-2011-ITN).
\end{acknowledgements} 

\appendix
\section{Bootstrap applied to transition matrices}
\label{app:A}

This appendix is concerned with the limited length of the record used to estimate the transition matrices $\{\hat{P}_\tau\}$ defined in section \ref{sec:estim}. This statistical inference problem can result in errors, notably in the rates and the regimes calculated from $\{\hat{P}_\tau\}$ in section \ref{sec:spectra} and \ref{sec:metastable} respectively, and requires an estimation of confidence intervals.

Following \citet{Chekroun2014}, we use a version of the non-parametric bootstrap \citep{Efron1982a, Mudelsee2010}, adapted to the estimation of transition matrices as done in \citep{Craig2002a}. It starts from the matrix $T_\tau$ counting the transitions of $\{y_t\}$ from one grid box to another, before the normalization has been applied to get the transition probabilities. From $T_\tau$, $N_s$ surrogate count matrices $\{T_{\tau, s}^*\}_{1 \le s \le N_s}$ are generated. This is done, for each row $i$, by taking with replacement $n_i$ target grid boxes among the $n_i$ transitions starting from grid-box $B_i$ of $T_\tau$. This accounts for drawing $n_i$ times from a Multinomial distribution with vector of probabilities $\{(\hat{P}_\tau)_{il}\}_{1 \le l \le m}$. From these $N_s$ surrogate count matrices, the transition matrices $\{\hat{P}_{\tau, s}^*\}_{1 \le s \le N_s}$ are then calculated by normalizing their rows. These matrices can then be used to estimate the sampling error of any functions of the transition matrix $\hat{P}_\tau$.

A thousand surrogate matrices $\{\hat{P}_{\tau, s}^*\}_{1 \le s \le N_s}$ are used to compute $99\%$ confidence intervals for the rates represented in figure \ref{fig:spectra}. For each lag $\tau$, the leading rates have been calculated for every surrogate transition matrix in $\{\hat{P}_{\tau, s}^*\}_{1 \le s \le N_s}$. For each rate $r_i(\tau)$ one gets a distribution of surrogate rates $\{(r_i(\tau))_s^*\}_{1 \le s \le N_s}$. After sorting these distributions, the $(0.005*N_s)^{th}$ and the $(0.995*N_s)^{th}$ values are taken to give the lower bound and the upper bound, respectively, of the confidence interval for rate $r_i(\tau)$.

In this study, we added a bias correction to the intervals because the described construction of the surrogate matrices introduces a bias towards lower values for rates, in particular for the secondary rates. We give an example to explain this bias. Assume that only two transitions start from a box $B_i$, with targets $B_k$ and $B_l$. In the surrogate, the probability to pick two different transitions ($B_k$ and $B_l$) will be of $1 / 2$, while the probability to pick two of the same transitions (twice $B_k$ or twice $B_l$) will also be of $1/2$. Thus, a bias is introduced towards a weaker mixing resulting in lower values of the rates (in particular to the secondary ones more sensitive to such details). This bias was removed by centering the mean of the surrogate rates $\{(r_i(\tau))_s^*\}_{1 \le s \le N_s}$ to the the value of the original rate $r_i(\tau)$ being tested.

\begin{figure}
	\centering
	\includegraphics[width=8.5cm]{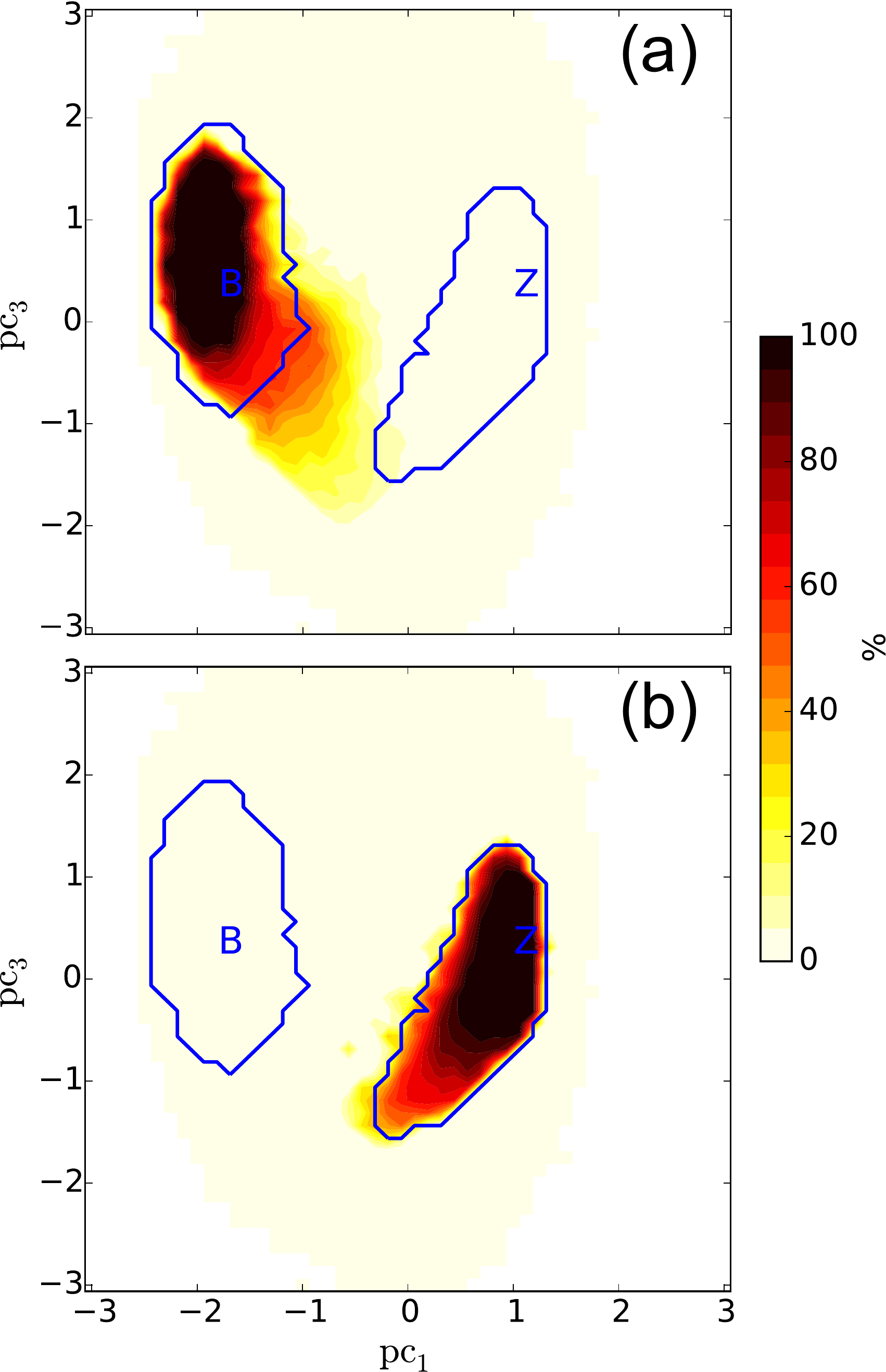}
	\caption{The value at each grid box represents the fraction of surrogate regimes for which the grid box belongs to (a) the blocked regime or (b) the zonal regime out of a $100$ surrogates. The surrogate regimes were detected by applying the algorithm of section \ref{sec:metastable} to the bootstrap surrogate transition matrices.}
	\label{fig:BSReg}
\end{figure}

The surrogate transition matrices $\{\hat{P}_{\tau, s}^*\}_{1 \le s \le N_s}$ can also be used to test the robustness of the regimes to the sampling by running, for each $\hat{P}_{\tau, s}^*$, the regime detection algorithm described in section \ref{sec:metastable}. The membership of a grid box to a regime can be given by a membership matrix $M_{i \beta}$ such that $M_{i\beta} = 1$ if grid box $B_i$ belongs to regime $E_\beta$, and is zero otherwise. Confidence intervals are not well suited for statistics taking logical values. Instead, we compute, for each grid-box $B_i$, the fraction of surrogates for which $B_i$ belongs to regime $E_\beta$. These fractions are plotted in figure \ref{fig:BSReg}a,b for the blocked and the zonal regime,  respectively,  using $N_s = 100$ surrogates.

We can see that the core of the regimes are very robust to the sampling, as indicated by the amount of grid boxes in each regime for which a fraction close to a $100\%$ of the boxes of the surrogates have been attributed to the same regime. However, boxes along the path of the transition from the zonal to the blocking regimes are less robust to the sampling. Such exit-region and entering-region for the zonal and the blocked regime, respectively, are indeed harder to attribute to a regime, since they correspond to regions where the invariance is weak. A possibility would be to add a simulated annealing step in the almost-invariant sets detection algorithm as has been used, for example, in \citet{Rosvall2008}.

\section{Robustness to grid resolution}
\label{app:B}

The robustness of the rates and the regimes to the grid resolution for which the transition matrices are estimated is tested next. 

\begin{figure}
	\centering
	\includegraphics[width=8.5cm]{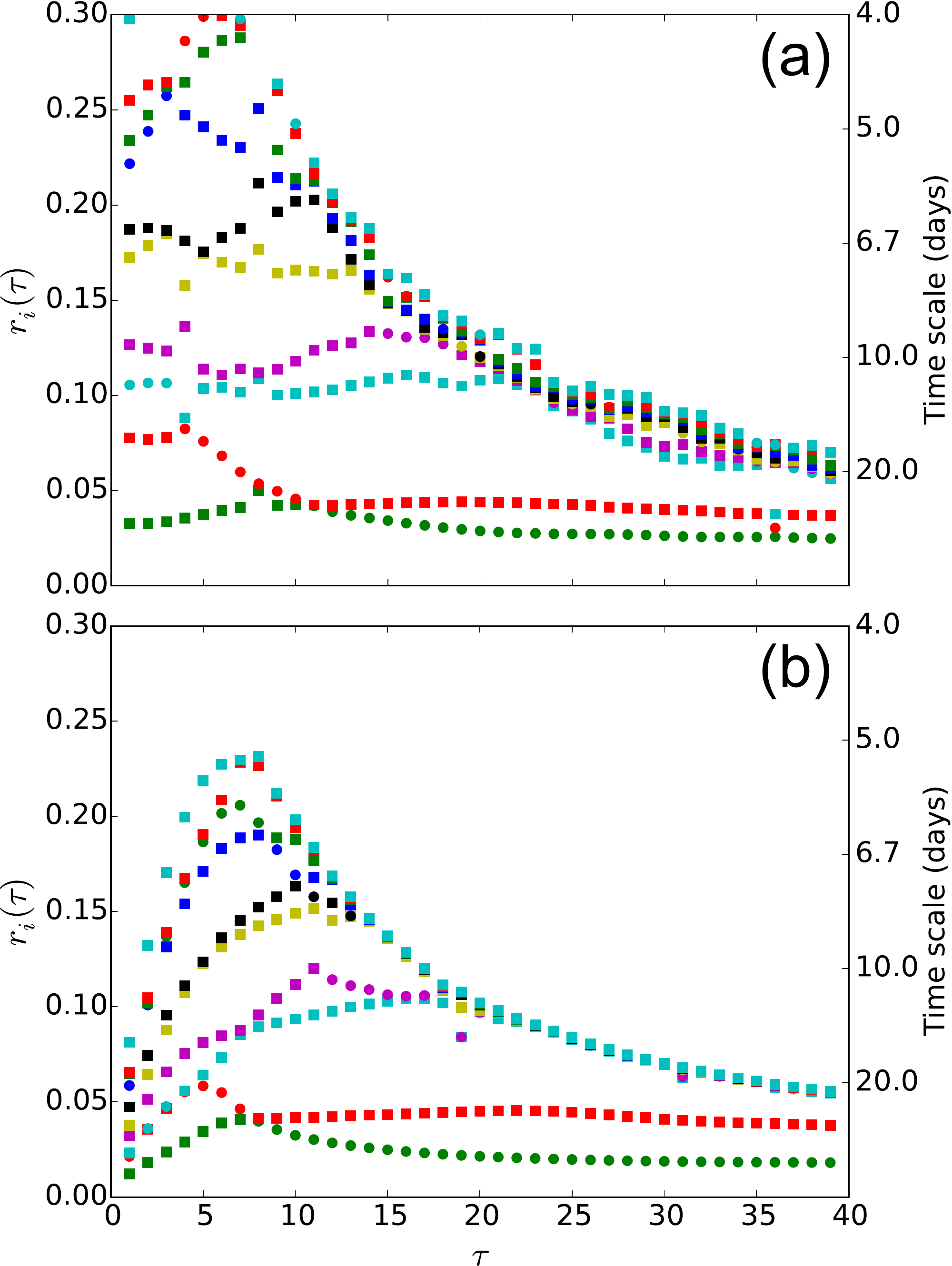}
	\caption{Rates $r_i(\tau)$ as in figure \ref{fig:spectra} but calculated from estimates of transition matrices on a (a) $10\times10$ grid and (b) $100\times100$ grid.}
	\label{fig:GridSpectra}
\end{figure}

Figure \ref{fig:GridSpectra}a, b represent rates calculated from transition matrices estimated on a grid of $10\times10$ and $100\times100$ boxes respectively, to be compared with figure \ref{fig:spectra} of section \ref{sec:spectra}. We can see that at least the $3$ leading rates in green, red and cyan are not much affected by the grid resolution so that the analysis of section \ref{sec:spectra} remains valid under changes of the grid in this range. This robustness to the grid of the leading rates is in agreement with the fact that they represent slow large-scale motions less likely to be affected by small perturbations of the transition matrices.

\begin{figure}
	\centering
	\includegraphics[width=8.5cm]{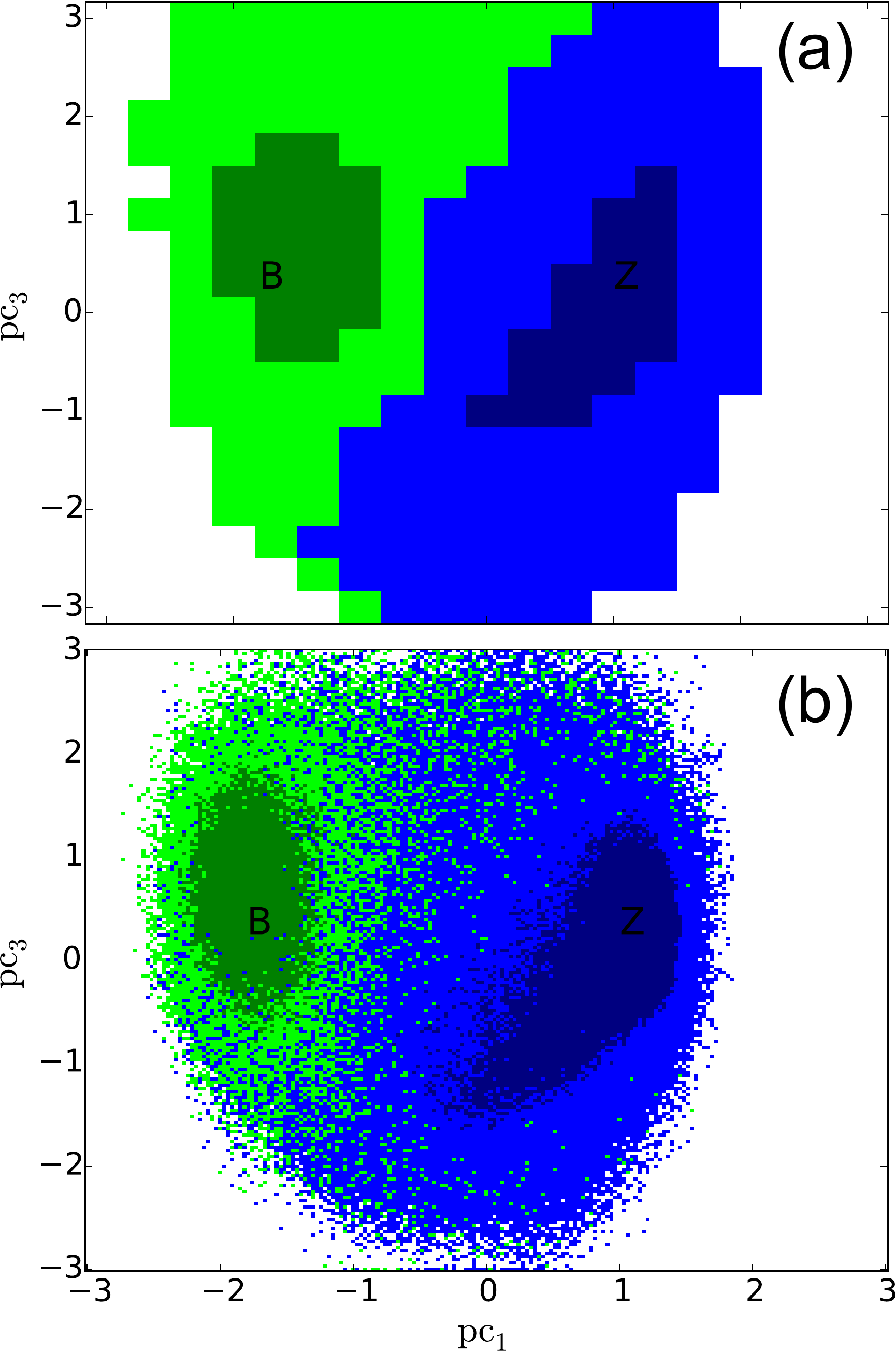}
	\caption{Almost-invariant sets and their regimes as for figure \ref{fig:regimes} but for a grid of (a) $20\times20$ and (b) $200\times 200$.}
	\label{fig:GridRegime}
\end{figure}

Figure \ref{fig:GridRegime}a,b represent regimes detected from transition matrices estimated on a grid of $20\times20$ and $200\times200$ boxes respectively, to be compared with figure \ref{fig:regimes} of section \ref{sec:metastable}. It can be seen that in both cases the regimes are very much alike to the one plotted figure \ref{fig:regimes} for the grid of $50\times50$ boxes. The quality of the regimes deteriorates for grids coarser than $20\times20$ (not represented here) but remains relatively good for resolutions as refined as $200\times200$ (Fig.~\ref{fig:GridRegime}.b). This robustness of the regimes to very refined resolutions can be explained by the aggregative nature of the almost-invariant sets detection algorithm. Indeed, one expects the estimates of the transition probabilities between grid boxes to deteriorate as the grid becomes thinner and as the number of samples by grid box decreases. However, because the algorithm iteratively agglomerates grid boxes into clusters, the transition probabilities between these coarser and coarser clusters become less and less sensitive to the sampling as the number of samples by clusters increases with their size.

\section{Robustness of the regimes to the time lag}
\label{app:C}

\begin{figure}
	\centering
	\includegraphics[width=8.5cm]{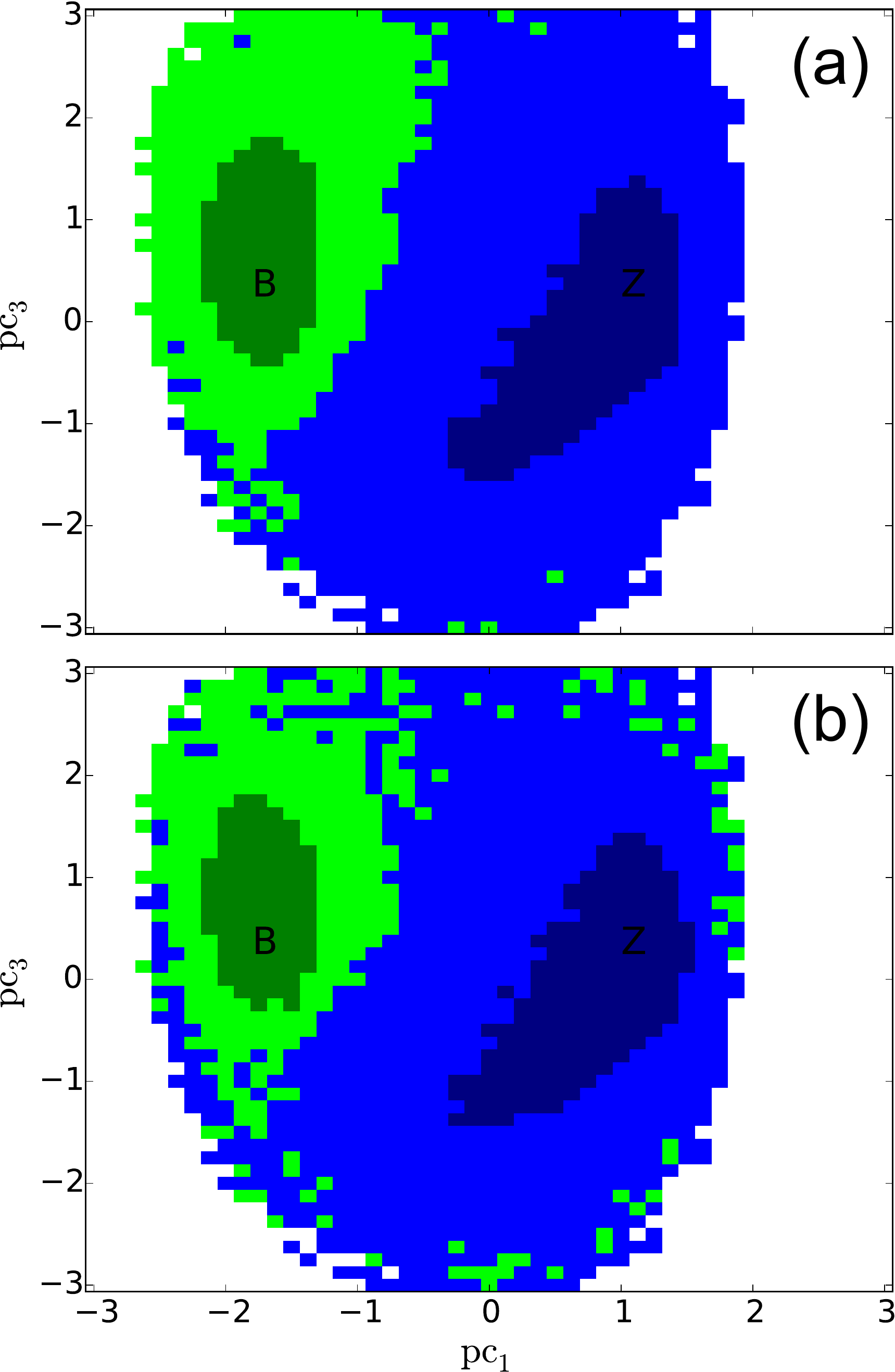}
	\caption{Almost-invariant sets and their regimes as for figure \ref{fig:regimes} but for a lag of (a) 7 days and (b) 50 days.}
	\label{fig:LagRegime}
\end{figure}

In addition  to the robustness test of the regimes to the grid resolution, we next test their 
robustness to the choice of the time lag of the transition matrix from which they are detected. Regimes detected for lags $\tau$ of 7 and 50 days are plotted in figure \ref{fig:LagRegime}a,b,  respectively, to be compared to figure \ref{fig:regimes} of section \ref{sec:metastable}. In this range, we can see that the quality of the regimes is robust to the lag.

For lags shorter than 7 days, however, the algorithm is not able to distinguish the transition path between the zonal and the blocked regime from the regimes themselves (not shown here). Such result is in agreement with the fact that for lags smaller than $7$ days, the complex pair of rates represented by green squares in figure \ref{fig:spectra}, more likely to be associated with the transition, is dominant over the real eigenvalue represented by red circles and likely to be associated with the meta-stability of the regimes. Thus the motions associated with the transition appear to be dominant over the meta-stability of the regimes for such short lags.

The deterioration of the regimes for lags larger that $50$ days is to be expected due to the fact that, as seen section \ref{sec:spectra}, the leading rates of figure \ref{fig:spectra}, associated with meta-stability, have time-scales not larger than $40$ days, so that for longer lags, the motions associated with the meta-stable regimes are likely to decorrelate and cannot be detected by the algorithm.

\section{Test of the semigroup property}
\label{app:D}

In this appendix, we justify that the distances plotted in figure \ref{fig:semigroup}a in section \ref{sec:spectra}, between densities transferred by a power of 2 of $\hat{P}_8$ and densities transferred by $\hat{P}_{2\times8}$, are mostly explained by memory effects rather than by the Galerkin approximation or by the sampling.

\begin{figure}
	\centering
	\includegraphics[width=8.5cm]{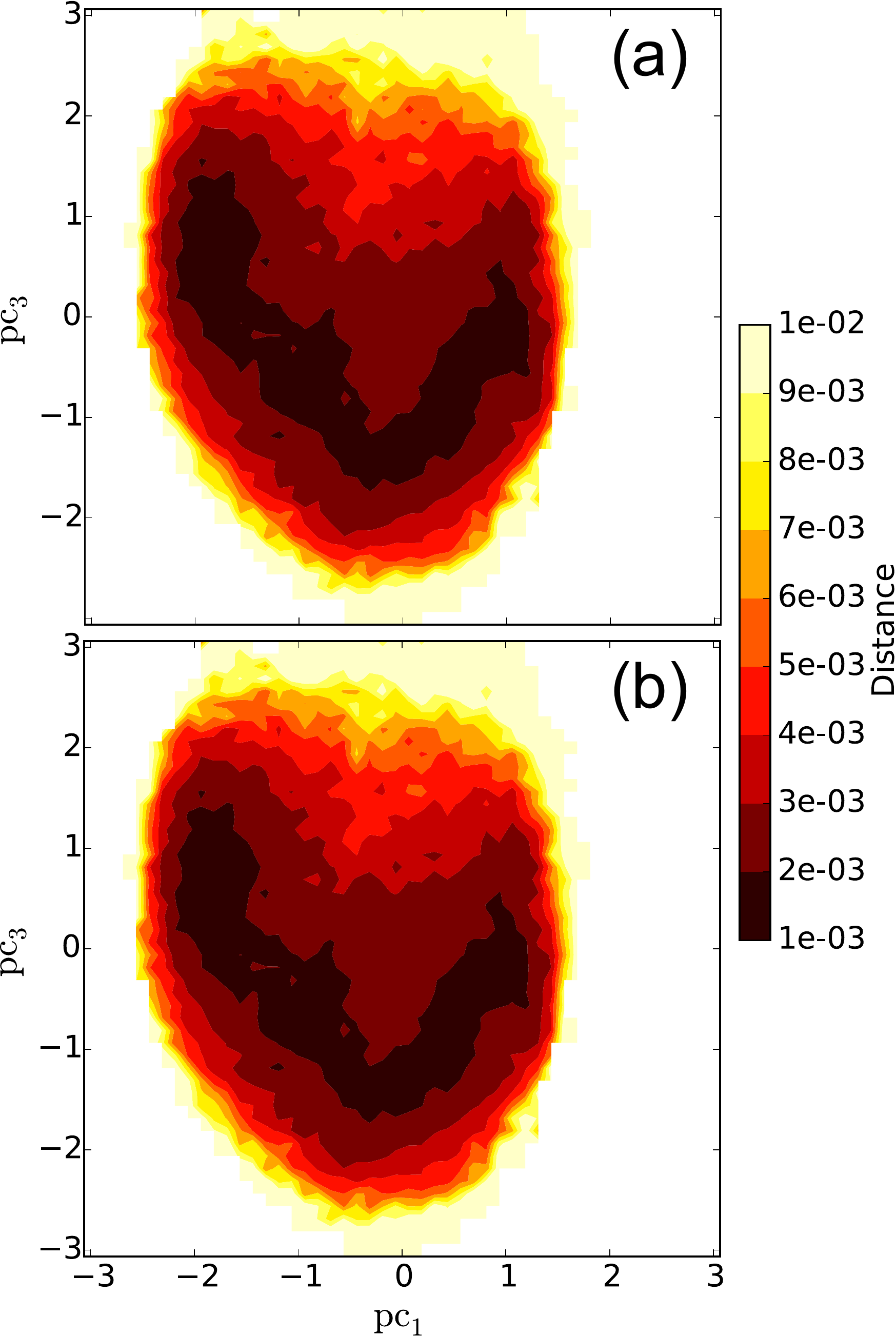}
	\caption{As for figure \ref{fig:semigroup}, the value at each grid-point represents the distance, for an initial density of 1 at this grid-point, between the density $f_{8k}^\mathrm{pow}$ transferred by the $k^{th}$ poser of $\hat{P_8}$ and a density $f_{8k}^\mathrm{long}$ transferred by $\hat{P}_{k\times 8}$ for $k = 2$. However, the transition matrices have been estimated from a time series of (a) $250,000$ days and (b) $1,000,000$ days, compared to the $500,000$ of figure \ref{fig:semigroup}.}
	\label{fig:semigroupLength}
\end{figure}

To do so, we first represent the same distances but calculated from transition matrices estimated from time series of only $250,000$ realizations and as much as $1,000,000$ realizations (Fig.~\ref{fig:semigroupLength}.a and b respectively), compared to the $500,000$ realizations from which the distances represented figure \ref{fig:semigroup} have been calculated. We can see that neither plots seem to be affected by the short or long length of the time-series. One has to go to record lengths shorter than $200,000$ realizations to start to see a difference in the distances (not shown here). Thus, the distances represented in  figure \ref{fig:semigroup} are not likely to be due to the limited length of the time-series from which the transition probabilities have been calculated.

\begin{figure}
	\centering
	\includegraphics[width=8.5cm]{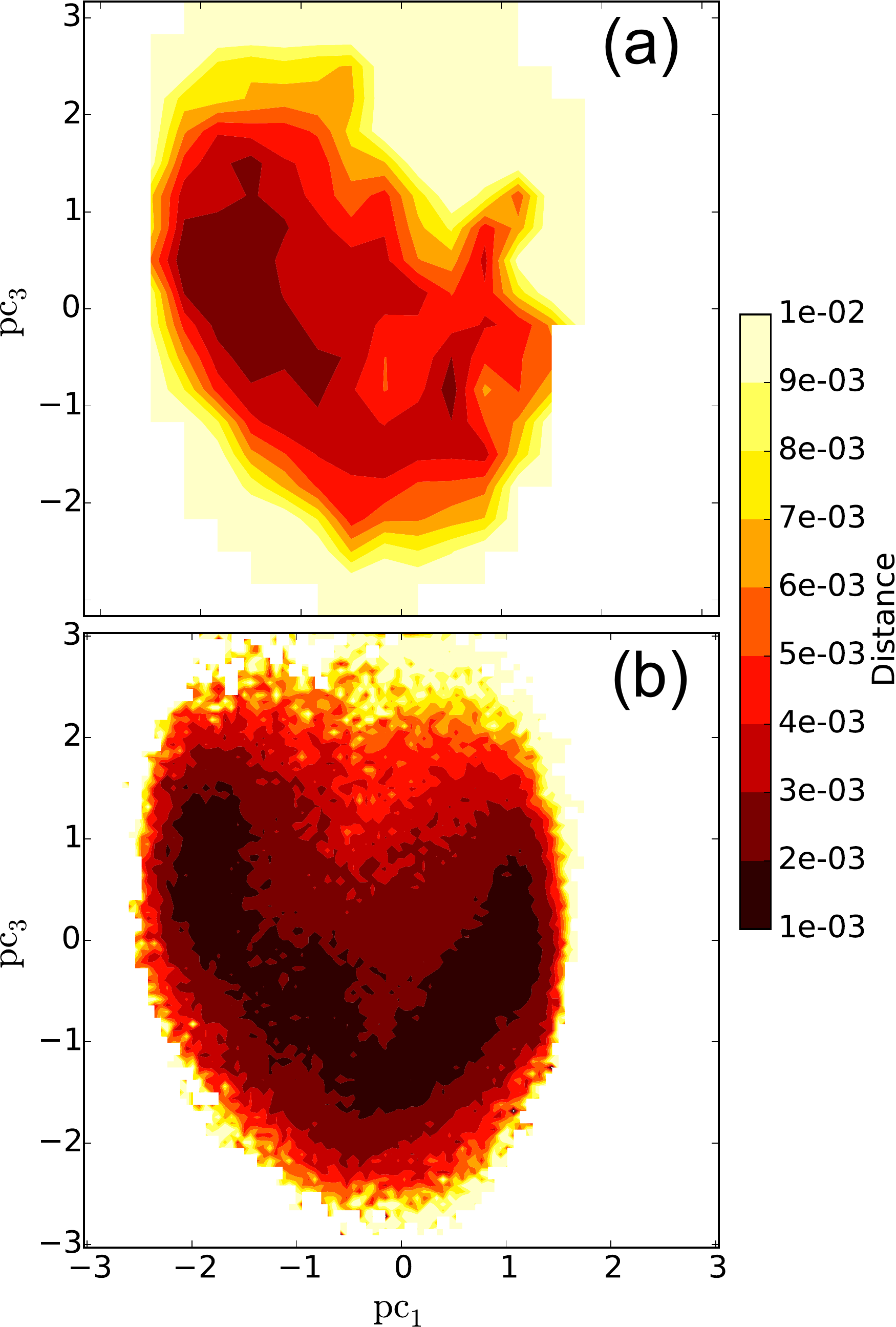}
	\caption{As for figure \ref{fig:semigroup}, the value at each grid-point represents the distance, for an initial density of 1 at this grid-point, between the density $f_{8k}^\mathrm{pow}$ transferred by the $k^{th}$ poser of $\hat{P_8}$ and a density $f_{8k}^\mathrm{long}$ transferred by $\hat{P}_{k\times 8}$ for $k = 2$. However, the transition matrices have been estimated from on a grid of (a) $20\times20$ and (b) $100\times100$ boxes, compared to the $50\times50$ boxes of figure \ref{fig:semigroup}.}
	\label{fig:semigroupGrid}
\end{figure}

Secondly, to test the effect of the Galerkin approximation on the distances, we have again reproduced figure \ref{fig:semigroup}, using $500,000$ realizations, but with a grid of $20\times20$ and $100\times100$ (Fig.~\ref{fig:semigroupGrid}.a and b respectively). While for the $20\times20$ grid, the distances get larger, indicating that this increase is due to the Galerking approximation, the distances for the $100\times100$ grid resolution (Fig.~\ref{fig:semigroupGrid}.b) are very much alike the one for the $50\times50$ resolution (Fig.~\ref{fig:semigroup}). Note that this is also true when a record length of $1,000,000$ realizations (not shown here), so that the similarities between the plots is not likely to be due to a compensating effect between the grid resolution and the number of samples per grid point.

We can thus conclude that the distances represented in figure \ref{fig:semigroup} for $\hat{P}_8^2$ and $\hat{P}_{2\times8}$ are mostly due to memory effects introduced by the only partial observation of the barotropic model.

%
%
%
%

\newpage 
\section*{References} 
\bibliography{chaos_reply1}

\end{document}